\documentclass[superscriptaddress,amsmath,amssymb, aps, pra,twocolumn]{revtex4-2}
\usepackage[T1]{fontenc}
\usepackage[latin9]{luainputenc}
\usepackage{amsmath}
\usepackage{amssymb}
\usepackage{graphicx}
\usepackage{float}
\usepackage{xcolor}
\usepackage[colorlinks]{hyperref}
\makeatletter
\usepackage{verbatim}

\makeatother

\newcommand{\orcid}[1]{\href{https://orcid.org/#1}{\includegraphics[width=7pt]{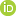}}}

\newcommand{\ket}[1]{\mbox{$ | #1 \rangle $}}
\newcommand{\bra}[1]{\mbox{$ \langle #1 | $}}

\makeatletter

\usepackage{babel}

\begin{document}

\title{Exploring quantum thermodynamics with NMR}

\author{C. H. S. Vieira\orcid{0000-0001-7809-6215}}
\email{carloshsv09@gmail.com}
\affiliation{Center for Natural and Human Sciences, Federal University of ABC, Avenida dos Estados 5001, 09210-580, Santo Andr\'{e}, S\~{a}o Paulo, Brazil.}

\author{J. L. D. de Oliveira\orcid{0000-0003-3743-4841}}
\affiliation{Center for Natural and Human Sciences, Federal University of ABC, Avenida dos Estados 5001, 09210-580, Santo Andr\'{e}, S\~{a}o Paulo, Brazil.}

\author{J. F. G. Santos\orcid{0000-0001-9377-6526}}
\affiliation{Faculdade de Ci\^{e}ncias Exatas e Tecnologia, Universidade Federal da Grande Dourados,
Caixa Postal 364, Dourados, CEP 79804-970, MS, Brazil.}
\affiliation{Center for Natural and Human Sciences, Federal University of ABC, Avenida dos Estados 5001, 09210-580, Santo Andr\'{e}, S\~{a}o Paulo, Brazil.}

\author{P. R. Dieguez\orcid{0000-0002-8286-2645}}
\affiliation{International Centre for Theory of Quantum Technologies (ICTQT), University of Gda\'{n}sk, Jana Bazynskiego 8, 80-309 Gda\'{n}sk, Poland.}

\author{R. M. Serra\orcid{0000-0001-9490-3697}}
\email{serra@ufabc.edu.br}
\affiliation{Center for Natural and Human Sciences, Federal University of ABC, Avenida dos Estados 5001, 09210-580, Santo Andr\'{e}, S\~{a}o Paulo, Brazil.}
\affiliation{Department of Physics, Zhejiang Normal University, Jinhua 321004, China}

\begin{abstract}
Quantum thermodynamics seeks to extend non-equilibrium stochastic thermodynamics to small quantum systems where non-classical features are essential to its description. Such a research area has recently provided meaningful theoretical and experimental advances by exploring the wealth and the power of quantum features along with informational aspects of a system's thermodynamics. The relevance of such investigations is related to the fact that quantum technological devices are currently at the forefront of science and engineering applications. This short review article provides an overview of some concepts in quantum thermodynamics highlighting test-of-principles experiments using nuclear magnetic resonance techniques.
\end{abstract}

\maketitle

\section{Introduction}

The last two decades advances in experimental techniques allowed the design of a wide set of quantum devices with different applications, such as quantum computations \cite{Arute2019, Gedik2015, Jones2001,Jones2011}, quantum sensing and quantum cryptography \cite{Riofrio2017, Aasi2013, Garbe2020}, among others. We may say that in quantum device applications, the role played by thermodynamics is important, which is relevant for the optimal performance search and the understanding of its constraints due to dissipation and reversibility. In general, quantum devices operate at the micro- and nano-scales, where quantum fluctuations become as relevant as thermal fluctuations and a proper description of the energy exchange is in order. Quantum thermodynamics \cite{Adesso2018, Deffner2019, Kosloff2013, Goold, Anders2016, Brandao2015, Xuereb2016} has been building, in the last years, to describe properly energy exchange at a quantum scale. Quantum fluctuation theorems allow a solid framework and establish limits on the non-equilibrium thermodynamics of quantum systems \cite{serra2014, Seifert2012, Dorner, Mazzola, Micadei2020a, Micadei2019, Micadei2021, Campisi2013, Tasaki, Camati2018b, Talkner, Esposito, campisi, Denzler2021, Jarzynski2004, sin11, cam15, Henao2018, cam14}. Furthermore, the use of quantum systems as working fluids in different quantum thermal devices is an interesting way to boost the performance of thermal cycles beyond its classical counterparts \cite{Scovil1959, Scully2003, deOliveira2021, Camati2019, Klatzow2019, Peterson2019, felce2020quantum, felce2021refrigeration, Talkner2017, Elouard2017, Brandner2015, Lin2021, Bresque2021, Campisi2019, Ding2018, Jordan2020, Mohammady2017, chand2017single, chand2018critical, ElouardPRL2018, anka2021measurement}. Another prominent feature of quantum thermodynamics is the inclusion of quantum information (coherence and non-classical correlations, for instance) as additional resources for thermodynamic tasks~\cite{Deffner2019, Goold}. 

Different experimental platforms have been used to investigate quantum thermodynamics aspects, for instance, trapped ions \cite{Hu2020, Ivanov2019, Rossnagel2016}, quantum circuit electrodynamics \cite{Pekola2015Nature,Anders2016,Cattaneo2021}, quantum optics \cite{Elouard, Zanin2019,Passos2019}, optomechanical systems \cite{Brunelli2015, Pigeon2015}, nuclear magnetic resonance (NMR) \cite{serra2014, Micadei2019, Camati2018b, Denzler2021, Micadei2021, lisboa2022experimental} etc. The last platform also had a prominent role in the development and demonstration of quantum information protocols~\cite{Chuang_1997,Cory_1997,Laflamme_1998,Chuang_2005}. The success in conducting quantum thermodynamics experiments in NMR platforms is well evidenced, for instance, in the implementation of a spin quantum heat engine \cite{Peterson2019, Pal2019, lisboa2022experimental} and by means of verification of quantum fluctuation relations for processes and cycles \cite{serra2014, Denzler2021, Micadei2021}. Moreover, NMR has proved to be an excellent setup to verify that quantum correlations \cite{Maziero2013} may revert the heat flow between two systems \cite{Micadei2019}. These few examples illustrate the usefulness of the NMR setup as a relevant platform for quantum thermodynamics experiments. 

Although this article is not complete or has extensive coverage, we hope to offer an overview of interesting NMR experiments that can be seen as an introduction to the field. We present in some detail the experimental implementation as well as the quantum physical aspects of the considered models. Sec.~\ref{Non-equilibirum energy fluctuations} is dedicated to covering NMR experiments involving quantum fluctuation theorems, as well as the connection between irreversibility and entropy production in quantum systems. In Sec.~\ref{Quantum thermal devices Jefferson} we present quantum thermal devices experiments carried out in NMR setups. In particular, we explore the efficient operation of a spin quantum heat engine, the distinct behavior of a refrigerator with indefinite causal order, the characterization of quantum engines powered by measurements, the charging protocols in quantum batteries, and the high precision measurements of quantum thermometers. Sec.~\ref{Information and thermodynamics} covers recent experimental advances using the NMR setup in the study of the role of information in thermodynamics protocols. Finally, Sec.~\ref{Conclusion and outlook} draws some conclusions and an outlook.

\section{Non-equilibrium energy fluctuations} \label{Non-equilibirum energy fluctuations}
\subsection{Work and heat in non-equilibrium quantum systems}
Energy changes characterized by heat and work have a central interest in thermodynamic processes. For a quantum system described by the Hamiltonian, $H_t$, and in the state (described by its density operator), $\rho_t$, at a given time, $t$, the system's average (internal) energy is given by $\left\langle \mathcal{U} \right\rangle = \text{tr}(\rho_t H_t)$. The system's energy may change in time due to a process controlled by some parameter, for instance, a modification of an external potential or due to the interaction with an environment, and both $H_t$ and $\rho_t$ may change in time from an initial configuration at $t=0$ to a final one at $t=\tau$. The average energy variation is computed as
\begin{equation}
\left\langle \Delta \mathcal{U} \right\rangle = \text{tr}(\rho_\tau H_\tau - \rho_0 H_0 ),
\end{equation} 
which comprises the two types of energy transfer, i.e., work and heat that depends on the process details (as in the first law of thermodynamics \cite{Callen1985}). As will be discussed in what follows, the separation of work and heat contributions to the internal energy change (as well as its experimental characterization), in a quantum system under general dynamics, is not an easy task~\cite{Adesso2018,Deffner2019,Goold,Anders2016,Myers2022}.

Let us begin with a simple scenario. Considering a unitary evolution of a system, as a nuclear spin under a time-modulated radio-frequency (rf) field on resonance. Formally, the system's density operator evolves as $\rho_t = U_t \rho_0 U^\dagger_t$, where $U=T_> \exp\left( -\frac{i}{\hbar}\int_0^\tau dt \, H_t \right)$ is the time evolution operator and $T_>$ is the time ordering operator. From an experimental point of view, we are considering here an evolution that avoids appreciable interaction of the system with its environment. In other words, all the dynamics that we are interested in happen at a time scale significantly smaller than the typical system's relaxation times~\cite{serra2014}. The energy variation, in this case, occurs in a controllable and useful way due to the change in the time-dependent Hamiltonian driven by an external potential. Thus, we can unambiguously identify the energy change as work, $\left\langle W \right\rangle = \left\langle \Delta \mathcal{U} \right\rangle$, in the unitary evolution of a quantum system. 

In a second simple scenario, let us consider a fixed system's Hamiltonian in time and after some initial state preparation, the system undergoes a non-unitary evolution due to its interaction with some environment. The system dynamics may be described by the Redfield formalism~\cite{Redfield1965}, a master equation \cite{{Petruccione2007,Celeri2011}} or a completely-positive trace-preserving (CPTP) map~\cite{Nielsen2011}, $\mathcal{M}(\rho) = \sum_j K_j \rho K^\dagger_j$, where $\{K_j\}$ are the Kraus operators describing the decoherence process. In this case, the energy transfer from or to the environment occurs in a sort of uncontrolled way accompanied by an entropy variation, and it is naturally identified as heat, $\left\langle Q \right\rangle = \left\langle \Delta \mathcal{U} \right\rangle$.

For general processes in an open quantum system that involves also time-dependent Hamiltonians, unequivocal definitions of heat and work are still a challenging problem in the field. We recall that these thermodynamic quantities are not directly observable or associated with Hermitian operators~\cite{Talkner}, they depend on the process details itself~\cite{Rubi2008, Aspuru-Guzik2017, Boukobza2019, Anders2018}. There are some efforts to overcome such challenges~\cite{Alicki1979,Parrondo,Deffner2015,Rezakhani2016,Fan2020,Weimer,Alonso2016,Kosloff2022}. Some of the available approaches are not fully consistent with each other and with the expected results in different scenarios. A framework adopted in some references is a formulation that depends on particular conditions such as a slow-driven Hamiltonian, resulting in dynamics without transitions between instantaneous energy eigenstates, strict energy conservation between subsystems, and a very weak coupling with a Markovian environment. In this particular scenario, with Gibbs thermal states, we can write, for instance, the time derivative of the mean internal energy as 
\begin{equation}\label{intern_ener_deriv}
\frac{d}{dt}\left\langle \mathcal{U} \right\rangle = \text{tr}(\dot{\rho_t} H_t)+ \text{tr}(\rho_t \dot{H_t}).
\end{equation}
From Eq.~(\ref{intern_ener_deriv}), conventionally, the mean heat absorbed by the system is identified as $\left\langle Q \right\rangle = \int_0^\tau dt \, \text{tr}(\dot{\rho_t} H_t)$ and the mean work performed on the system as $\left\langle W \right\rangle = \int_0^\tau dt \, \text{tr}(\rho_t \dot{H_t})$~\cite{Anders2016, Alicki1979, Nori2007}. Similar to the classical (conventional) thermodynamics, work and heat are not state variables and depend on how the process is carried out while the average internal energy variation, $\left\langle \Delta \mathcal{U} \right\rangle$ only depends on the start and end points.

The later work and heat definitions inspired in Eq.~(\ref{intern_ener_deriv}) are not suitable when non-Gibssian states are present in the dynamics or in a scenario where a time-dependent Hamiltonian induces transitions between instantaneous energy eigenstates, e.g., a fast drive such that the Hamiltonian does not commute in different instants of time $t$ and $t'$, $[H_t,H_{t'}]\neq0$. In this case, the unitary (driven Hamiltonian) part of the dynamics may introduce coherence in the system's state which can be misidentified as heat, since this change in the system's state (energy eigenstates transitions due to fast dynamics) is not associated with entropy variation. Therefore, besides other features, coherence in quantum thermodynamics may introduce ambiguity in the work and heat definitions. Recently, interesting and useful approaches to unambiguous work and heat definitions were put forward explicitly associating the internal energy change with the entropy variation~\cite{Ala-Nissila2022, Bernardo2020}. In this case, the contribution to the internal energy change causing variation in von-Neumann entropy is then identified as heat, on the other hand, the contribution not related to entropy variation is identified as work~\cite{Ala-Nissila2022, Bernardo2020}. The work and head definitions in quantum thermodynamics are subtle and should be carefully addressed. In the next sections, we will explore the experimental characterization of energy fluctuations, work, and heat in different applications using NMR.

\subsection{Quantum fluctuation relations}
Analogous to the context of classical stochastic systems~\cite{Seifert2012}, in the quantum regime, the concepts of work and heat are associated with statistical expectation values~\cite{Adesso2018, Deffner2019}. An important framework in quantum thermodynamics is provided by fluctuation relations~\cite{Jarzynski,Crooks,Tasaki,Talkner,Esposito,campisi}, in particular, the Crooks theorem (detailed form)~\cite{Crooks,Tasaki} and Jarzynski identity (integral form)~\cite{Jarzynski}. Considering explicitly energy fluctuations in non-equilibrium dynamics, such relations connect equilibrium properties of thermodynamic relevance with non-equilibrium features. Fluctuation relations are among only a few equalities known to be valid beyond the linear-response regime~\cite{Seifert2012,Talkner,Esposito,campisi} and can be regarded as out-of-equilibrium generalizations of the second law, with applications also to unconventional scenarios, where access and manipulation of microscopic system's configurations are present~\cite{Camati2018b}. To investigate fluctuation theorems it is necessary to measure and characterizes energy fluctuations in processes involving work and/or heat. Besides the conceptual questions associated with work and heat characterization (as discussed in the previous section), it is also necessary to measure transition probabilities among energy eigenstates along an out-of-equilibrium process. This is, in general, a challenging task due to the measurement precision required. 

In 2014, Batalh\~{a}o et al.~\cite{serra2014} performed the first experimental verification of a detailed and integral version of fluctuation relations for work in a quantum system, using the excellent control present in NMR. The experiment reported in Ref.~\cite{serra2014} was carried out using liquid-state NMR spectroscopy of the $^1$H and $^{13}$C nuclear spins of a $^{13}$C labeled chloroform sample (CHCl$_3$) diluted in deuterated acetone on a Varian 500 MHz Spectrometer. 

Initially, the authors used spatial averaging methods (rf-pulses, free evolution under the scalar, and gradient pulses)~\cite{key-22} to prepare the $^{1}$H-$^{13}$C nuclear-spin pair in the pseudo-thermal state equivalent to $\rho_{HC}^0=\ket{0}\bra{0}_H \otimes\rho_C^0$, with $\rho_C^0=e^{-\beta H_0}/Z_0$ a Gibbs thermal state of the $^{13}$C nuclear spin at effective spin temperature $T$. Throughout this article, we will use the qubit representation of a spin-1/2 with $\{\ket{0},\ket{1}\}$ corresponding to the ground and excited states, respectively. We also used in the previous expressions, the inverse temperature $\beta= (k_B T)^{-1}$ (where $k_B$ is the Boltzmann constant), the partition function $Z_0=\mathrm{tr}(e^{-\beta H_0})$, and an effective initial $^{13}$C Hamiltonian $H_0 = h\nu_1\sigma_y^C$ set by a traverse rf field. The sample used in Ref.~\cite{serra2014} can be regarded as an ensemble of identical, non-interacting, spin-$1/2$ pairs due to its small concentration.

\begin{figure}[ht]
\includegraphics[width=0.97\columnwidth]{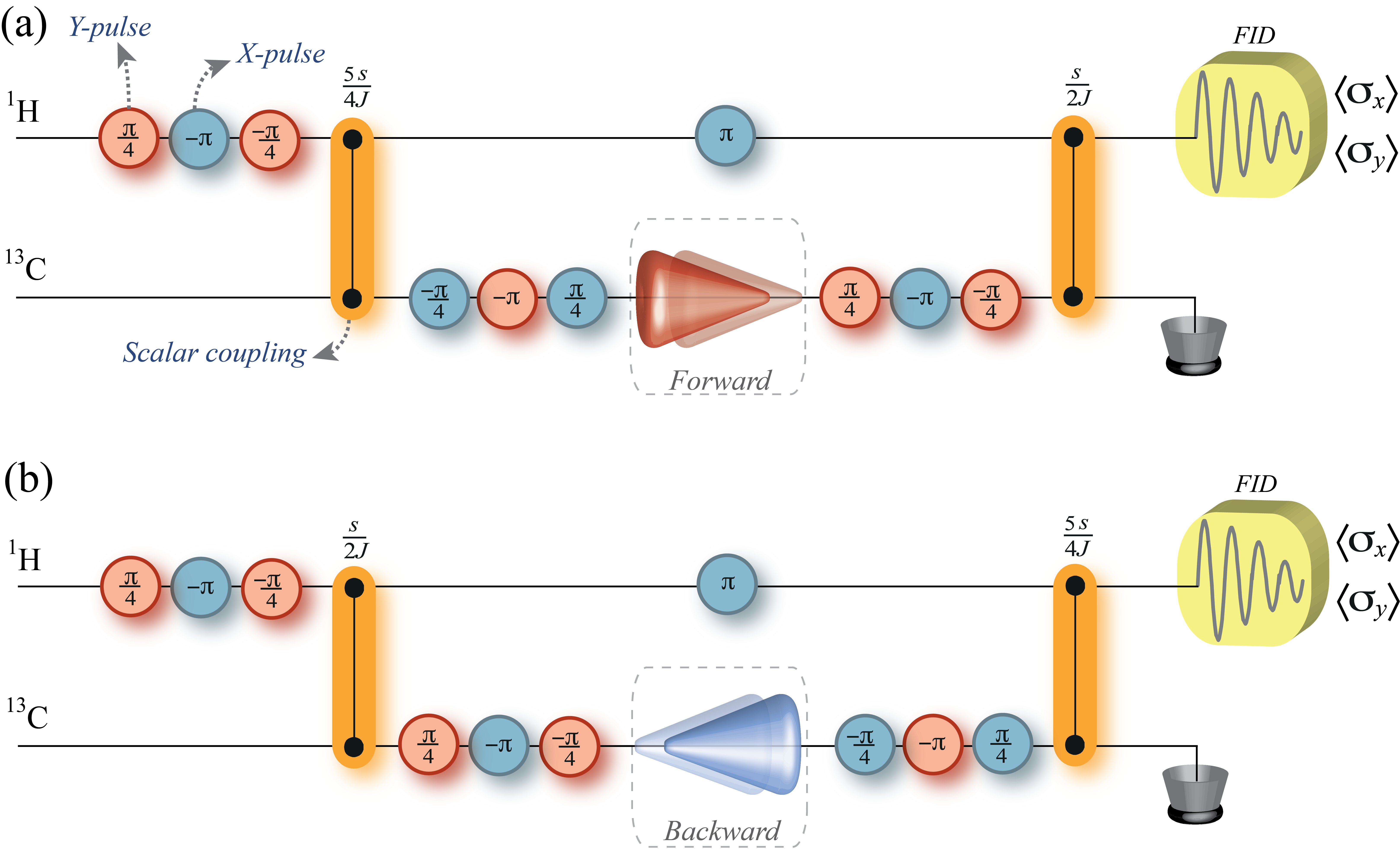}
\caption{NMR pulse-sequence for interferometric measurement of the work probability distribution. (a) Pulse sequence for the measurement of the work characteristic function $\chi(u)$ in the forward unitary process. (b) Sequence for the measurement of $\chi(u)$ in the backward process. The blue (red) circles represent transverse rf-pulses with phases and amplitudes adjusted to produce a $x$ ($y$) rotation by the displayed angle. Free evolution under the scalar coupling $ H_J=\frac{\pi\hbar}{4} J \sigma^H_z \sigma^C_z$ (with $J \approx 215.1$~Hz) are represented by orange junctions. The time-length of the scalar coupling is set by the parameter $s$, which is related to the conjugate variable $u$ in Eq.~(\ref{eq:workcharfunc}) by $s = 2\pi \nu_1 u $. Figure adapted from Ref.~\cite{serra2014}.}
\label{fig:circuit}
\end{figure} 

In the experiment reported in Ref.~\cite{serra2014} the $^{13}$C nuclear spin is the driven system, while the proton ($^{1}$H) is an ancillary qubit used to investigate energy fluctuations in the first one. After the initial state preparation a time-modulated rf pulse on resonance with $^{13}$C nuclear spin was applied, to produce effectively the following time-dependent Hamiltonian (in the double rotating frame with respect to the $^{1}$H and $^{13}$C resonance frequency):
\begin{align}
 H_t = h \nu_t 
 \left( \sigma_x^{C}\sin\frac{\pi t}{2\tau}  + \sigma_y^{C}\cos\frac{\pi t}{2\tau}  \right),
 \label{eq:quench}
\end{align} 
where $\sigma^C_{x,y,z}$ are the Pauli matrices for the $^{13}$C nuclear spin and $\nu_t=\nu_1 \left(1-t/{\tau}\right) + \nu_2 t/{\tau}$ is a linear ramp (taking an overall time $\tau= 0.1$~ms) of the rf field with the intensity adjusted such that $\nu_1 = 2.5\text{ kHz}$ to $\nu_2 = 1.0\text{ kHz}$, $t \in [0,\tau]$.

\begin{figure*}[ht]
\includegraphics[width=0.95\textwidth]{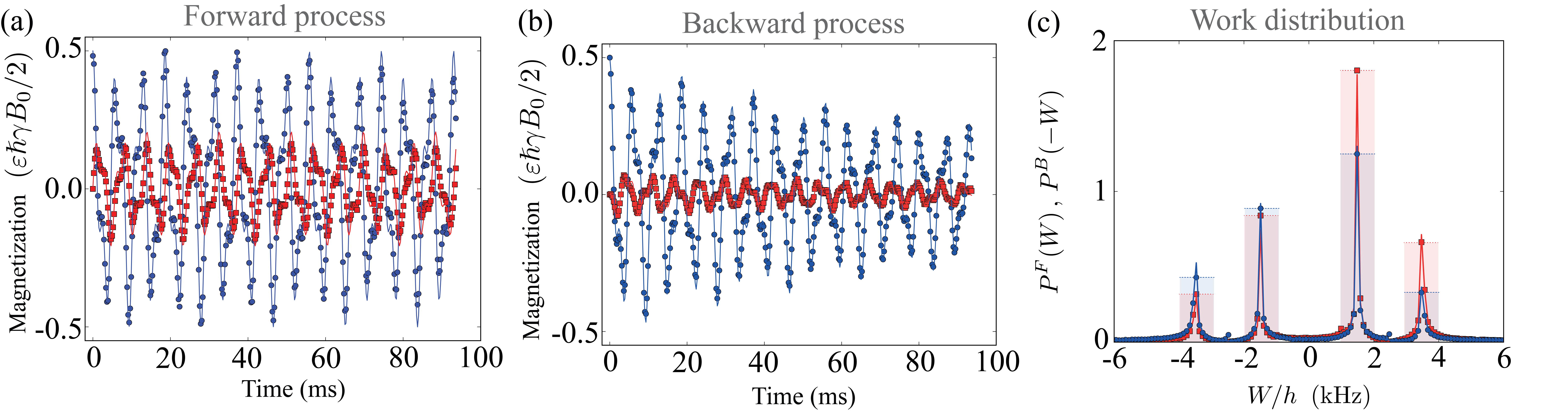}
\caption{Experimental characteristic function and work distribution for the forward and backward processes. (a) The blue circles (red squares) show the normalized experimental data for the $x$-component ($y$-component) of the $^1$H transverse magnetization at the effective spin temperature $k_bT/h = 6.9 \pm 0.3$~kHz, representing the work characteristic function for the forward process. (b) work characteristic function for the backward process. The solid lines show Fourier fittings, which are in excellent agreement with the theoretical simulation of the process. The horizontal axis is the evolution time for the shortest coupling, $s/(2J)$ in the pulse sequence implemented for the reconstruction of $\chi_F(u)$ (in Fig.~\ref{fig:circuit}). The uncertainty in the temperature is due to finite precision in the initial state preparation. The parameter $\varepsilon$ is the ratio between magnetic and thermal energies, $\gamma$ is the gyromagnetic ratio, and $B_0 \approx 11.75$~Tesla is the longitudinal magnetic field intensity (along the $z$-direction) used in the experiment~\cite{serra2014}. (c) Displays the modulus of the inverse Fourier transform of the transverse magnetization representing the work distribution. Red squares (blue circles) are the experimental points for the forward (backward) process. The horizontal axis was inverted for the backward process. The experimental data are well fitted by a sum of four Lorentzian peaks centered at $\pm1.5\pm0.1$~kHz and $\pm3.5 \pm 0.1$~kHz (solid lines), in agreement with the theoretical expectation that predicts the location of the peaks to be at $\pm (\nu_1\pm\nu_2)$. The amplitudes of the peaks, from the leftmost to the rightmost in each panel, are proportional to the probabilities $p^{0}_1 p^\tau_{0|1}$, $p^{0}_1 p^\tau_{1|1}$, $p^{0}_0 p^\tau_{0|0}$, $p^{0}_0 p^\tau_{1|0}$ respectively. Figure adapted from Ref.~\cite{serra2014}.}
 \label{fig:workdist}
\end{figure*} 

For such a unitary driven Hamiltonian, the mean work performed by the rf field on the $^{13}$C nuclear spin can be written as $\langle W\rangle=\int P(W) \,dW$, where the work probability distribution is 
\begin{equation}
P(W)=\sum_{n,m}p_{n}^{0}p^\tau_{m|n}\delta\left(W-\epsilon_{m}^\tau+\epsilon_{n}^0\right),
\label{eq:workdist}
\end{equation}
with $p_{n}^{0}=e^{-\beta \varepsilon_n^0}/Z_0$ being the probability to find the system in the initial energy eigenstate $|n(0)\rangle$ (with
energy { $\epsilon_{n}^0$}), the transition probability $p^\tau_{m|n}=\left|\langle m(\tau)|U_{\tau}|n(0)\rangle\right|^{2}$
is the conditional probability of driving the system to the instantaneous
energy eigenstate $|m(\tau)\rangle$ (with energy { $\epsilon_{m}^\tau$})
at the end of the evolution, given the initial state $|n(0)\rangle$, where $U_{\tau}$ is the time evolution operator. The process can be regarded as a unitary evolution since the driving time is much smaller than the relaxation time scales $T_1$ and $T_2$ (of an order of seconds) for this sample~\cite{serra2014}. The time-modulated Hamiltonian in Eq.~(\ref{eq:quench}) can lead the system to an out-of-equilibrium state, in the sense that, after the evolution, it can not be written as a Gibbs distribution at the effective inverse spin temperature $\beta$ for $H_\tau$. A backward (reversal) version of the protocol~\cite{Camati2018b} was also implemented preparing $^{13}$C nuclear spin in a Gibbs thermal state of the $H_\tau$, $\rho_C=e^{-\beta H_\tau}/Z_\tau$, and performing reverse driving with $H^{B}_t=-H_{\tau-t}$. The negative signal emulates the Hermitian conjugation of the evolution operator $U_t^\dagger$~\cite{Camati2018b}.

To investigate energy fluctuations, the authors of Ref.~\cite{serra2014} used a Ramsey-like interferometric scheme~\cite{Mazzola, Dorner, Campisi2013} put forward by the pulse sequence displayed in Figs.~\ref{fig:circuit}(a) and (b). The characteristic function of the work distribution 
 \begin{equation}
\label{eq:workcharfunc}
\begin{aligned}
\chi(u)&=\int{P_{F}(W)e^{-iuW}}dW \\
&=\sum_{m,n}p^0_{n}p^\tau_{m\mid n}e^{iu({\epsilon}_{m}-\epsilon_{n})}, 
\end{aligned}
\end{equation}
for the carbon nuclear spin was then encoded in the transverse magnetization of the final proton state as $\operatorname{Re}[\chi(u)]=2 \langle \sigma^H_x \rangle$ and $\operatorname{Im}[\chi(u)]=2 \langle \sigma^H_y \rangle$~\cite{serra2014}. We note that in the protocol sketched in Figs.~\ref{fig:circuit}(a) and (b), the scalar coupling between the two nuclei, $H_J$, is present during the whole evolution. The $x(y)$-rotations were produced with square pulses with a time duration of about $10$~$\mu$s in Ref.~\cite{serra2014}, so during these operations, the effect of the scalar coupling can be neglected in a first approximation (due to the value of the frequency $J$). Along the driving evolution (of order of $100$~$\mu$s) of the $^{13}$C nuclear spin, Eq.~(\ref{eq:quench}), the scalar coupling could have a mild effect on the dynamics, which is effectively mitigated by the $\pi$ rotation on the proton in the middle of the driving protocol. The orange connections in Figs.~\ref{fig:circuit}(a) and (b) represent free evolution under the scalar coupling between the two nuclei, which are, in fact, a delay time along the pulse sequence. An interesting point for further developments of this technique, for energy-fluctuation spectroscopy, might be the incorporation of decoupling schemes on the proton nuclei along the $^{13}$C driving evolution, in to improve precision.

The work distribution of the driving process (the time modeled rf field) was obtained from the inverse Fourier transform of the measured $\chi(u)$. For a spin-1/2 system, it is possible to observe, in an out-of-equilibrium fast dynamics, four transitions between the initial ($t=0$) and final states ($t=\tau$), as for instance, ground to ground ($n=m=0$), ground to excited ($n=0$ and $m=1$) and so on. In this way, four well-defined peaks were observed in $P(W)$ for the forward and backward process~\cite{serra2014} as reproduced in Fig.~\ref{fig:workdist}(c), associated with one of the four possible transitions ($n \rightarrow m$ with $n,m=0,1$). The transition probabilities, $p^\tau_{m|n}$, are independent of the initial state effective temperature and they are fixed only by the driving Hamiltonian [Eq.~(\ref{eq:quench})]. In the reported experiment it was determined as $p^\tau_{1|1}= 0.71 \pm 0.01$, $p^\tau_{0|0}\approx 0.69\pm0.01$, and $p^\tau_{0|1}\approx p^\tau_{1|0}\approx 0.71 \pm 0.01$ for both forward and backward processes. It can also be understood as a verification of the micro-reversibility hypothesis~\cite{serra2014}. As expected, the measured work distribution $P(W)$ is described by Lorentzian functions, due to the resolution of the finite time Fourier analyses, instead of delta functions present in the theoretical expression in Eq.~(\ref{eq:workdist}). We note that the observation time of the work characteristic function [Figs.~\ref{fig:workdist}(a) and \ref{fig:workdist}(b)] is ultimately limited by the relaxation time scales $T_1$ and $T_2$. 

The experimentally reconstructed work distributions for forward and backward processes can be used to verify detailed and integral fluctuation relations \cite{Jarzynski,Crooks,Tasaki,Talkner,Esposito,campisi} in a genuinely quantum scenario, associated with driving Hamiltonians that do not commute in different times. The result of Ref.~\cite{serra2014} represented an important step towards the assessment of out-of-equilibrium dynamics in quantum systems. 

From a theoretical point of view, the energy fluctuations in unitary driving should satisfy a detailed fluctuation relation written as \cite{Crooks,Tasaki}
\begin{equation}
\label{TS}
\frac{P^F(W)}{P^B(-W)}=e^{\beta(W-\Delta F)},
\end{equation}
where $\Delta F$ is the net change in the free energy due to the driving process and is theoretically written in terms of initial and final partition function as $\beta\Delta F=-\ln(Z_\tau/Z_0)$. 

\begin{figure}[ht]
\includegraphics[width=0.95\columnwidth]{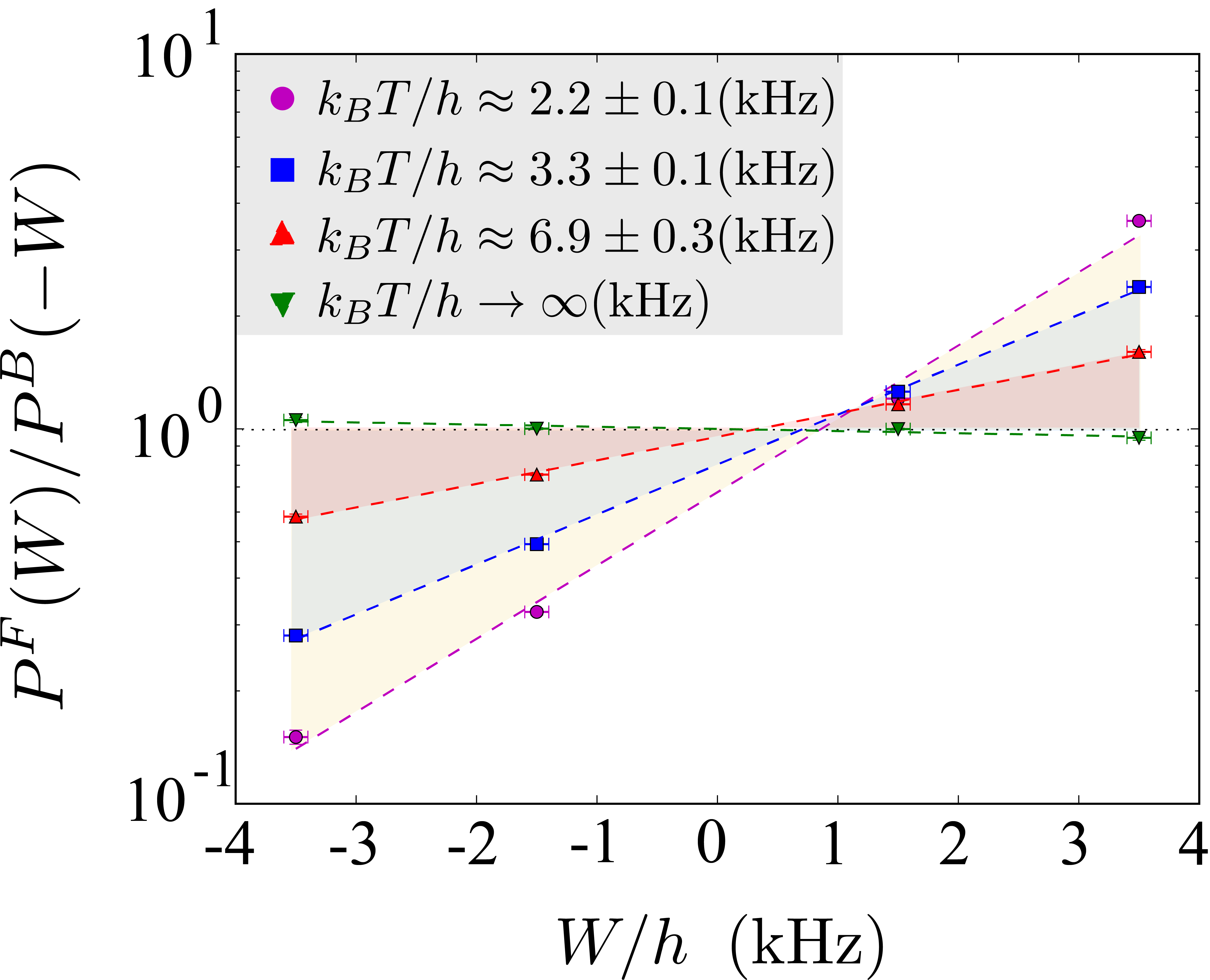}
\caption{Detailed fluctuation relation for work in a unitary driving. The ratio $P^F(W)/P^B(-W)$ is plotted in the logarithm scale for the four peak frequencies in the work distribution. From a linear fit of the data, we can obtain the free energy variation $\Delta F$ and the inverse of the temperature $\beta$. Figure adapted from Ref.~\cite{serra2014}.}
\label{fig:TS}
\end{figure} 

One of the most important uses of Eq.~(\ref{TS}) is for an experimental determination of free energy variations in a given process. Using the measured work distribution for different temperatures, the authors of Ref.~\cite{serra2014} verified a detailed fluctuation relation in a quantum system and used it to experimentally determine a change in free energy. Figure~\ref{fig:TS} displays the left-hand side of Eq.~\eqref{TS} in a linear-logarithmic scale for the different initial effective spin temperatures under the unitary driving described above. The experimental points are in good agreement with the expected linear relation, $\ln(P^F(W)/P^B(-W))={\beta(W-\Delta F)}$, thus confirming the predictions of the Crooks theorem [Eq.~\eqref{TS}]. The data can also be used as a high-precision out-of-equilibrium thermometer (the curve slop in Fig.~\ref{fig:TS}) which can capture tiny temperature variations ($\simeq 5\, \pm \,5\%$~nK in Ref.~\cite{serra2014}). The horizontal error bars shown in Fig.~\ref{fig:TS} are associated with the Fourier spectral linewidth, which depends on the number of oscillations resolved in the work characteristics function (Fig.~\ref{fig:workdist}).

An integral version of the fluctuation relation for work (the Jarzynski identity) was also verified in Ref.~\cite{serra2014} at a quantum regime. Although it is a corollary of the integral fluctuation relation, the Jarzynski identity was previously formulated~\cite{Jarzynski} as
\begin{equation}
\label{Jarid}
\langle e^{-\beta W}\rangle=e^{-\beta\Delta F}
\end{equation}
where the average is taken over the forward process $P_F(W)$. It can be seen as a nonequilibrium generalization of the Clausius statement for the second law, $\langle W\rangle -\Delta F \geq 0$, obtained by applying Jensen's inequality. To verify the relation in Eq.~(\ref{Jarid}), the authors of Ref.~\cite{serra2014} used the experimentally determined value of free energy variation, $\Delta F$, obtained from the data in Fig.~\ref{fig:TS} or through the relation $\langle e^{-\beta W} \rangle = \chi(i\beta)$, resulting from an analytical continuation of the characteristic function. In Ref.~\cite{serra2014} it was also observed a very good agreement between the experimental data and theoretical predictions for the Jarzynski identity. We can say that this contribution opened the possibility of performing energy fluctuation spectroscopy in out-of-equilibrium thermodynamics of quantum systems taking advantage of the excellent control and precise magnetization measurements available in NMR.

In Ref.~\cite{Pal2019} the quantum heat-exchange fluctuation relation \cite{Jarzynski2004} was verified through $^{19}$F NMR spectroscopy in a 1,1,2-trifluoro-2-iodoethane sample, employing the interferometric method from \cite{serra2014} and the formalization scheme introduced in \cite{Peterson2019}. Also making use of similar methods, Peterson et al. \cite{Peterson2019} experimentally studied the marginal distribution for work, $P(W)$ and heat $P(Q)$ in a quantum Otto cycle employing a spin-1/2 system as a working substance, that will be described with some details in Sec. \ref{sec:quantum Otto engines}. 

The joint distribution for work and heat $P(W,Q)$ was explored in Ref.~\cite{Denzler2021} to describe efficiency fluctuations in a quantum thermodynamic cycle, which allowed for the verification of a detailed fluctuation relation for the joint work and heat variables. For a quantum heat engine cycle (the quantum Otto cycle), a detailed fluctuation relation was theoretically obtained as \citep{sin11,lah12,cam14,cam15}
\begin{equation}
\frac{P(W,Q)}{P(-W,-Q)}=e^{\Delta\beta Q-\beta_{1}W},\label{eq:Fluc_rel}
\end{equation}
where $\Delta\beta=\beta_{1}-\beta_{2}$ are the difference between cold, $\beta_1$, and hot, $\beta_2$, reservoirs inverse temperature, $P(W,Q)$ is the joint work and heat
distribution of the quantum engine, and $P(-W,-Q)$ is the joint distribution of measuring $(-W,-Q)$ in the reverse cycle operation. An integral fluctuation theorem, 
\begin{equation}
\langle e^{-\Sigma}\rangle= \iint dWdQ~P(W,Q) e^{-\Sigma}=1,
\end{equation}
for the entropy production $\Sigma = \Delta\beta Q-\beta_{1}W$ can be obtained after integration over one cycle \cite{sin11,lah12,cam14,cam15}. The latter expression represents a nonequilibrium generalization of the Carnot formula, $\langle W\rangle /\langle Q\rangle \leq 1-T_1/T_2$, which can be derived by applying Jensen's inequality \cite{sin11,lah12,cam14,cam15}. 

\begin{figure}[ht]
\includegraphics[width=0.95\columnwidth]{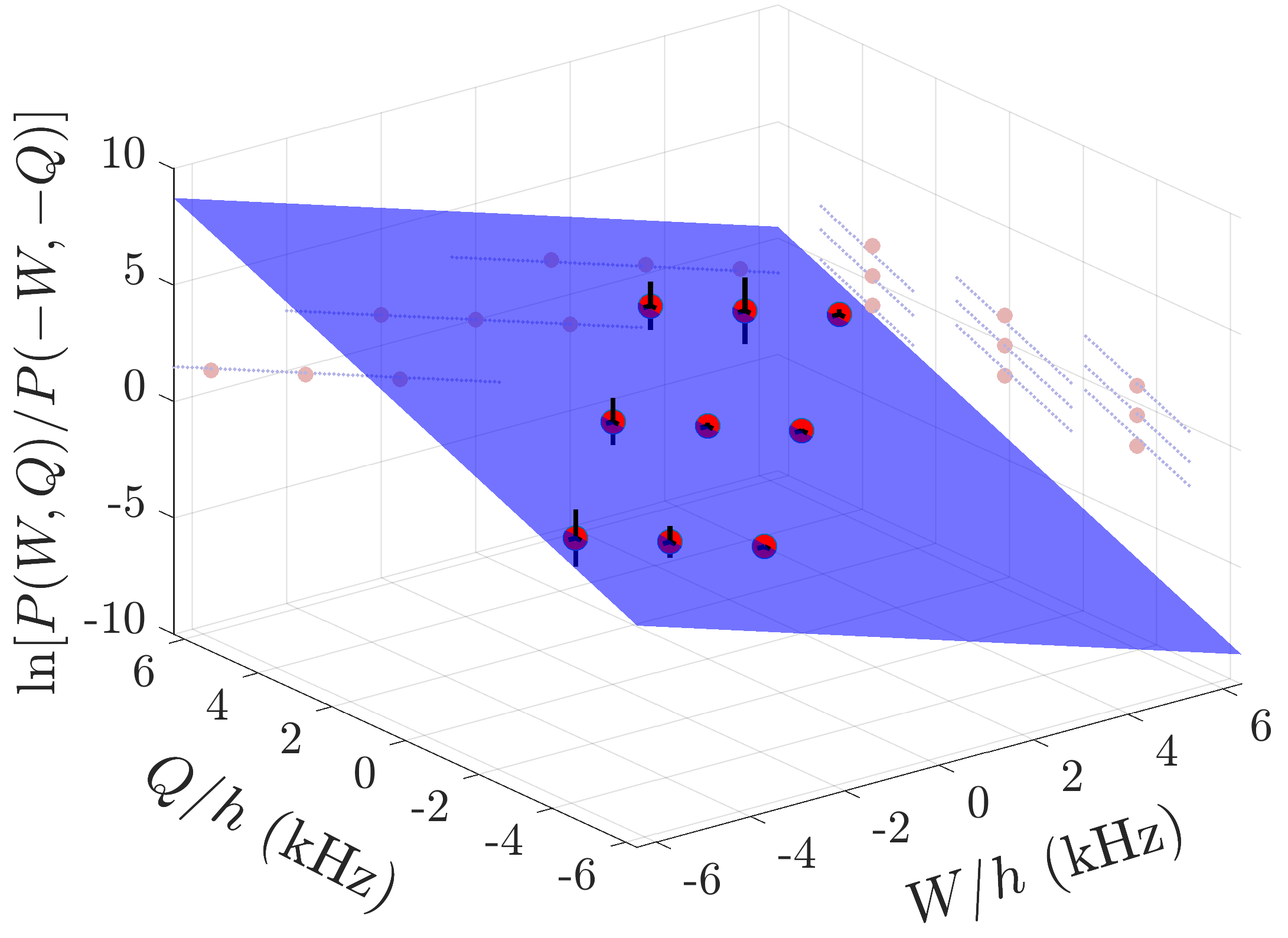}
\caption{Experimental verification of the detailed fluctuation relation for work and heat in a quantum Otto cycle. The values of $\ln\left[P(W,Q)/P(-W,-Q)\right]$ (red dots) should lie within the (blue) plane defined by the total entropy production $z= \Sigma =\Delta\beta Q-\beta_{1}W$; the dashed blue lines show the respective projections of the plane on the work and heat axes. Figure adapted from Ref.~\cite{Denzler2021}.}
\label{fig:jointfluct}
\end{figure}

Denzler et al.~\cite{Denzler2021} using an experimentally reconstructed work and heat joint distribution, $P(W,Q)$, for a quantum Otto cycle on a $^{13}$C nuclear spin as a working substance in a chloroform sample, verified the prediction of Eq.~(\ref{eq:Fluc_rel}). Fig.~\ref{fig:jointfluct} displays such verification for a cycle with driving time of $\tau=200$ $\mu$s performed in Ref.~\cite{Denzler2021}.
We can observe a very good agreement between the experimental values of $\ln\left[P(W,Q)/P(-W,-Q)\right]$ (red dots) and
the predictions of Eq.~(\ref{eq:Fluc_rel}) indicated by the (blue) plane $z=\Sigma$, where $z$ is the vertical axis. 

A heat exchange fluctuation theorem in the presence of quantum correlations among the systems in thermal contact was introduced in Ref.~\cite{Micadei2020a}. Such a theorem was experimentally verified by Micadei et al.~\cite{Micadei2021} applying dynamic Bayesian networks in NMR. Non-classical correlation has a nontrivial role \cite{Micadei2019,Henao2018} in heat transport, it can boost the heat transfer in the conventional direction, i.e., from hot to cold \cite{Henao2018}, or it can be consumed to revert usual flux \cite{Micadei2019}. We will explore this subject later on (Sec.~\ref{sec:thermodynamic_arrow_of_time}). 

Herrera et al.~\cite{Herrera2021} introduced a convenient method to obtain the work distribution in many body systems inspired by ideas from density functional theory (DFT). The method was successfully tested using NMR also in Ref.~\cite{Herrera2021}.

\subsection{Irreversibly, entropy production, and thermodynamic uncertainty relations}
The irreversible entropy production is a hallmark in out-of-equilibrium microscopic processes~\cite{Landi2021a}. First, let us consider a system initially in thermal equilibrium with an environment at inverse temperature $\beta$. Then the system undergoes a fast unitary driven evolution during a time interval $\tau$ followed by some energy dissipation to the environment. Here, we are again supposing that the driven evolution is fast enough to be considered unitary. In this scenario, the average entropy production $\langle \Sigma \rangle$ can be related with the Kullback-Leibler relative entropy between states $\rho^F_t$ and $\rho^B_{t-\tau}$ along the forward and backward (i.e. time-reversed) dynamics~\cite{Batalhao_2015,Kawai2007,Jarzynski_2009} as
\begin{equation}
\label{Entprod}
\langle \Sigma \rangle=\beta(\langle W \rangle -\Delta F) = \mathcal{S}_{KL}(\rho^F_t||\rho^B_{t-\tau}),
\end{equation}
where $\mathcal{S}_{KL}(\rho^F_t||\rho^B_{t-\tau})=\text{tr}[\rho^F_t(\ln\rho^F_t - \ln\rho^B_{t-\tau})]$. Equation~(\ref{Entprod}) quantifies irreversibility in terms of the microscopic quantum evolution. A
reversible process, $\langle \Sigma \rangle=0$, occurs if and only if the forward and backward microscopic dynamics are indistinguishable. The 
entropy production in a non-equilibrium quantum process and its fluctuations was measured using NMR in Ref.~\cite{Batalhao_2015}.

As commented before, fluctuations have a pivotal role in small quantum systems. In recent years, many efforts have been made to establish a trade-off between fluctuations and dissipation in stochastic thermodynamics. A set of relations addressing this goal
has been collectively referred to as thermodynamic uncertainty relations (TURs) \cite{key-1,key-2,key-3}. 

TURs dictate fundamental lower bounds for nonequilibrium currents, (i.e., work, heat exchange, etc.) in terms of total entropy production. In turn, denoting $\mathcal{Q}$ as a time-integrated current observable, a particular TUR imposes a bound as follows 
\begin{equation}
\frac{\mathrm{Var(\mathcal{Q})}}{\langle\mathcal{Q}\rangle{{}^2}}\geq\frac{2}{\langle\Sigma\rangle},\label{eq:1}
\end{equation}
where $\langle\mathcal{Q}\rangle$, $\mathrm{Var}(\mathcal{Q})=\langle\mathcal{Q}^{2}\rangle-\langle\mathcal{Q}\rangle^{2}$,
are the average and the variance of $\mathcal{Q}$, respectively,
and $\langle\Sigma\rangle$ is the total average entropy production. That seminal TUR, Eq.~(\ref{eq:1}), was initially discovered in biochemical networks, where the system is assumed to obey a Markovian continuous-time dynamic and begin from a nonequilibrium steady state \cite{key-1} being proved later by different approaches and techniques \cite{key-4,key-5,key-6}.
Subsequently, TURs also has been broadened for discrete-state Markov processes \cite{key-7}, periodically driven systems \cite{key-8,key-9}, multidimensional systems \cite{key-10}, first-passage times \cite{key-11,key-12}, measurement and feedback control \cite{key-13,key-14,key-15}, broken time-reversal symmetry systems \cite{key-16,key-17}, and for arbitrary initial states \cite{key-18}. In addition, stemming from fluctuations theorems (FT), generalized bounds have been derived for thermodynamic uncertainty relation, usually refereed as GTURs \cite{key-19,key-20}, which are expressed as follows
\begin{equation}
    \frac{\mathrm{Var}(\mathcal{\mathcal{\mathcal{Q}}})}{\langle \mathcal{Q}\rangle^{2}}\geq\frac{2}{e^{\langle\Sigma\rangle}-1},
\end{equation}
and
\begin{equation}
    \frac{\mathrm{Var}(\mathcal{\mathcal{\mathcal{Q}}})}{\langle \mathcal{Q}\rangle^{2}}\geq f(\langle\Sigma\rangle),
\end{equation}
where $f(x)=\cosh^{2}(g(x/2))$, and $g(x)$ is the inverse function of $x\tanh(x)$.

In the context of nuclear magnetic resonance (NMR), a recent experimental study of TUR and GTURs has been reported \cite{key-21}. In this work, the authors employed a liquid sample of sodium fluorophosphate and a 500 $\mathrm{MHz}$ Bruker NMR spectrometer to characterize the heat exchange between the $\mathrm{^{19}F}$ and $\mathrm{^{31}P}$ nuclei of this molecule by means of the so-called XY mode. The sample was initially
prepared in pseudo-thermal states corresponding to different effective spin temperatures~\cite{key-22}. The system's Hamiltonian is composed mainly by the interaction of each nucleus with the external magnetic field ${H}^F_{Z}+{H}^P_{Z}$
and by the scalar coupling $
H_{\text{int}}=\frac{\hbar\pi}{4}J\sigma_{z}^{\mathrm{F}}\sigma_{z}^{\mathrm{P}}$, 
where $J=868\,\mathrm{Hz}$ is the coupling constant between the $\mathrm{^{19}F}$ and $\mathrm{^{31}P}$ nuclei. 

In Ref. \cite{key-21}, the authors used a rf pulse sequence 
and the scalar coupling to effectively evolve the two spins system under an effective $\sigma_x\sigma_y$ coupling Hamiltonian and establish heat exchange
between the $\mathrm{^{19}F}$ and $\mathrm{^{31}P}$ nuclei. In this case, the total Zeemann Hamiltonian commutes with
the interaction term, i.g. $\left[H_{\text{int}},H_{Z}^{F}+H_{Z}^{P}\right]=0$,
implying that the energy change for one qubit is exactly compensated
by the other qubit. It means that there is no energy cost involved
during the process of turning on or off the interaction between the
qubits \cite{Micadei2019} and therefore the average entropy production turns out to be
\begin{equation}
\langle\Sigma\rangle=(\beta_{F}-\beta_{P})\langle Q\rangle,\label{eq:3}
\end{equation}
with $\beta_\alpha=(k_{B}T_\alpha)^{-1}$, $T_\alpha$ is the spin temperature for each nucleus and $\langle Q\rangle$
is the average heat exchange during the interaction time \cite{key-21}.

By performing quantum state tomography (QST) \cite{key-22} the authors obtained the cumulants of the heat exchange distribution. The
expression for the ratio $\langle Q^{2}\rangle/\langle Q\rangle{{}^2}$
was obtained in Ref.~\cite{key-21} using a perturbative approach up to the quadratic order of the thermal affinity $\Delta\beta$ enabling to explore some regions where the conventional TUR in Eq.~(\ref{eq:1}) is violated as already theoretically predicted. The authors also verified experimentally the validity of the GTURs.


\section{Quantum thermal devices} \label{Quantum thermal devices Jefferson}

Despite the quantum thermodynamics has been developed in many different directions in the last decade, it originally started with proposals of quantum heat engines (QHE), more specifically, with a seminal article by Scovil and Schulz-DuBois dated 1959~\cite{Scovil1959}. The authors considered the simplest model of a MASER (microwave amplification by stimulated emission), a pumped three-level system, where they demonstrated theoretically that these three-level MASERs can be regarded as continuous heat engines with two field modes playing the role of heat reservoirs at different temperatures, \(T_c\) and \(T_h\).

More recently, it was observed that quantum resources such as coherence, non-classical correlations, and squeezing can be employed to obtain an advantage in some thermal devices when compared with their classical counterparts \cite{Abah2014,deOliveira2022,Camati2019,Klatzow2019,Brandner2015,Klatzow2019,Bresque2021,Klaers2017}.

\subsection{Quantum Otto engines } \label{sec:quantum Otto engines}

The advances in quantum-devices control allowed to design and experimentally test quantum thermal machines where some non-classical signatures are present. For single-qubit working substance, quantum coherence plays a fundamental role in thermodynamic features as, for instance, the entropy production along the cycle and the extracted work, being directly associated with the cycle performance \cite{Camati2019, Klatzow2019}. The experimental implementation, as well as the study of the performance of a spin quantum engine caring out a quantum Otto cycle, has been done in Ref.~\cite{Peterson2019} using an NMR platform.

\begin{figure}[h]
\centering
	\includegraphics[scale=0.07]{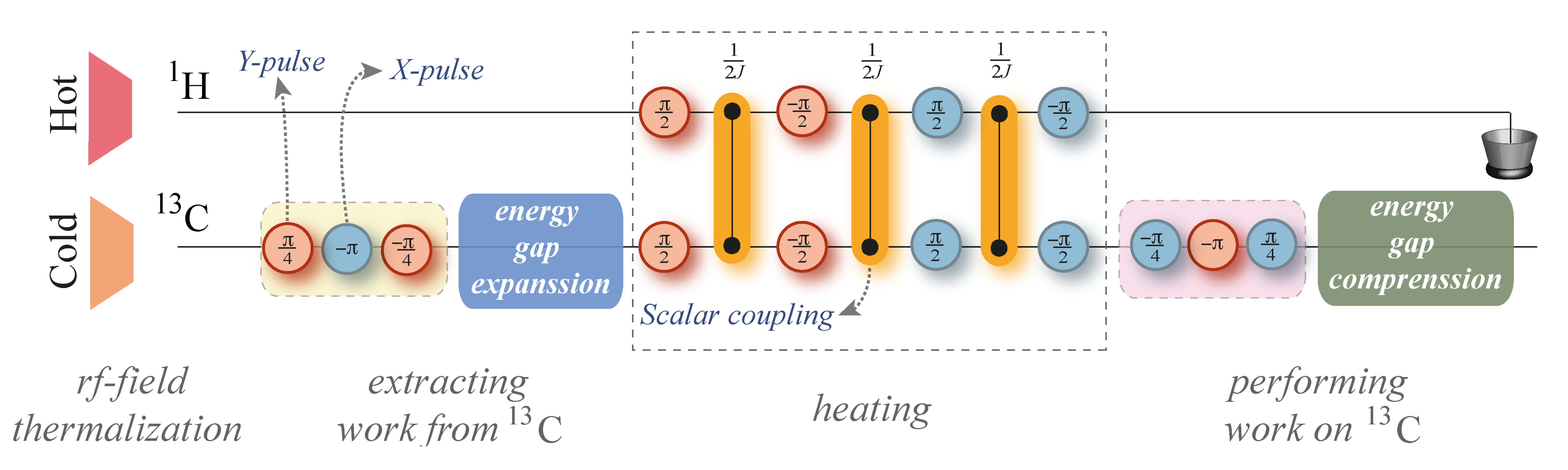}
	\caption{Simplified pulse sequence of the experimental protocol for the  quantum heat engine based in NMR. $\mathrm{^{1}H}$ and $\mathrm{^{13}C}$ nuclear spins are initially prepared in pseudo-thermal states corresponding to hot and cold spin temperatures, respectively. Blue (red) circles represent $x (y)$ rotations by the displayed angle produced by transverse rf pulses. Orange connections stand for free-evolution under the scalar interaction $(\mathcal{H}_J )$ during the time displayed above the symbol. The unitary driving for the energy gap expansion (compression) protocol is implemented by a time-modulated rf field resonant with the $\mathrm{^{13}C}$ nuclear spin. The Hydrogen nucleus is used to deliver the heat at the proper part of the cycle, working as a heat bus. Adapted from Ref.~\cite{Peterson2019}.}
	\label{cycle01}
\end{figure}

To perform the experiment reported in Ref.~\cite{Peterson2019}, it was employed a $\mathrm{^{13}C}$-labeled $\mathrm{CHCl_3}$ (Chloroform) liquid sample diluted in deuterated Acetone-d6 in a 500 MHz Varian NMR spectrometer. The spin \(1/2\) of the $\mathrm{^{13}C}$ nucleus was settled to be the working medium, while the \(\mathrm{^1H}\) nuclear spin was used as a \textit{heat bus} (the intermediate agent responsible for the heat exchange between the working medium and the hot reservoir). High radio-frequency (rf) modes near the hydrogen Larmor frequency (\(500\) MHz) played the role of the hot reservoir, while low rf modes near to carbon resonance frequency (\(125\) MHz) played the role of the cold reservoir.
The description of the engine performing the quantum Otto cycle is given by the following steps.

(I) Cooling process: Employing spatial average techniques generated by rf and gradient pulses, the $\mathrm{^{13}C}$ nuclear spin is initially prepared in a pseudo-thermal state, equivalent to $\rho_0^{eq,1} = e^{-\beta_1 {H}_1^C}/Z_1$ at a cold inverse spin temperature $\beta_1 = (k_B T_1)^{-1}$, with $Z_1 = \text{tr}\left[ e^{-\beta_1 H_1^{C}}\right]$ the corresponding partition function, $T_1$ is the absolute spin temperature of the cold reference state, and $H_1^{C}$ is the initial Hamiltonian of the working substance.

(II) Expansion process: The working substance is driven by a time-modulated rf field resonant with the $\mathrm{^{13}C}$ nuclear spin similar to Eq.~(\ref{eq:quench}). It is described initially by the Hamiltonian ${H}_1^{C} = -h\nu_1\sigma_x^C/2$, with $\nu_1 = 2.0$ kHz. The energy gap expansion happens in a driving time length much shorter than the typical decoherence timescales, which enables the description of this process as a unitary evolution, \(U_{\tau}\), that drives the $\mathrm{^{13}C}$ nuclear spin to an out-of-equilibrium state (\(\rho_{\tau}^{C}\)). The final Hamiltonian of the expansion protocol will be \(H_2^{C}=-h\nu_1\sigma_y^C/2\).

\begin{figure*}[ht]
\includegraphics[width=1\textwidth]{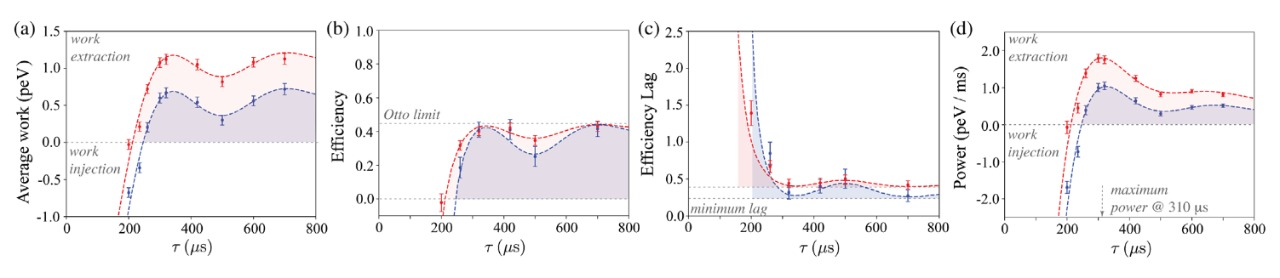}
\caption{Spin quantum engine figures of merit: (a) average extracted work, (b) efficiency, (c) efficiency lag due to entropy production, and (d) extracted power as a function of the driving protocol time length (\(\tau\)). Points represent experimental data. The dashed lines are based on theoretical predictions and numerical simulations. In all experiments, the spin temperature of the cold source is set at \(k_BT_1=(6.6\pm0.1)\) peV. Data in blue and red correspond to implementations with the hot source spin temperatures set at \(k_BT_2^A=(21.5\pm0.4)\) peV and \(k_BT_2^B=(40.5\pm3.7)\) peV, respectively. Adapted from Ref.~\cite{Peterson2019}.}
 \label{fig:QOE}
\end{figure*} 

(III) Heating process: The $\mathrm{^{13}C}$ nuclear spin absorbs heat from the $H_1^{C}$, which was prepared at a higher temperature until it reaches full thermalization at the hot inverse spin temperature \(\beta_2=(k_BT_2)^{-1}\). At the end of the process, the $\mathrm{^{13}C}$ nuclei are in the hot equilibrium states, equivalent to \(\rho_0^{eq,2}=e^{-\beta_2 H_2^C}/Z_2\). This is performed by the pulse sequence depicted in Fig.~\ref{cycle01}.

(IV) Compression process: Finally, following the time-reversed process of the expansion protocol, an energy gap compression is performed.

Despite the work extracted from (performed on) the $\mathrm{^{13}C}$ nuclear spin during the energy gap expansion (compression) driving protocol is a stochastic variable, described by a probability distribution \(P_{\text{exp}}(W)\) [\(P_{\text{comp}}(W)\)], the net extracted work from the engine can be assessed by the interferometric approach described in Ref.~\cite{serra2014}. In this experiment, the characteristic function \(\chi_\text{eng}(u)\) of the work probability distribution is measured, while the inverse Fourier transform of \(\chi_\text{eng}(u)\) provides the work probability distribution for the engine cycle \(P_{\text{eng}(W)}=\int \mathrm{d}u\chi_\text{eng}(u)e^{iuW}\). The average value of the extracted work can then be obtained by using \(\langle W_{\text{eng}}\rangle=\int\mathrm{d}WP_{\text{eng}}(W)W\). In contrast, the heat probability distribution \(P (Q)\) was determined by using two QSTs, so the mean heat from the hot source can be expressed as \(\langle Q_{\text{hot}}\rangle=\int\mathrm{d}Q P(Q)Q\).

The quantum-engine efficiency, $\eta = {\langle W_{\text{eng}} \rangle}/{\langle Q_{\text{hot}}\rangle}$, can be written in terms of efficiency lags (associated with entropy
production) as
\begin{equation}
\eta=\eta_{\text{Carnot}}-\mathcal{L},
\label{eq:efficience_lag}
\end{equation}
and the lag is given
by~\cite{Peterson2019} 
\begin{align}
\mathcal{L} & =\frac{\mathcal{S}_{KL}\left(\left.\rho^{\text{exp}}_{\tau}\right\Vert \rho_{0}^{eq,2}\right)+\mathcal{S}_{KL}\left(\left.\rho^{\text{comp}}_{\tau}\right\Vert \rho_{0}^{eq,1}\right)}{\beta_{1}\left\langle Q_{hot}\right\rangle },
\label{eq:lag}
\end{align}
where $\rho^{\text{exp}}_{\tau} = {U}_{\tau}\rho_{0}^{eq,1}{U}_{\tau}^{\dagger}$ is the state obtained after stroke II (gap expansion), $\rho^{\text{comp}}_{\tau} = {V}_{\tau}\rho_{0}^{eq,2}{V}_{\tau}^{\dagger} $ is the state obtained after stroke IV (gap compression), and $\eta_{\text{Carnot}}=1-\left.T_{1}\right/T_{2}$
is the standard Carnot efficiency.

Figure~\ref{fig:QOE}(a)-(d) illustrates the figure of merits obtained in the spin engine implemented in Ref.~\cite{Peterson2019}. It is possible to observe a lower bound on the time driving time length \(\tau\) (approximately at \(\approx 200\) \(\mu\)s), where the entropy production is so large [characterized by the efficiency lag, Fig.~\ref{fig:QOE}(c)] that it is not possible to extract work. Therefore, slower operation leads to better efficiency. However, in general, we are not interested in a too-slow engine with a small amount of power. The extracted power is maximized when the energy gap expansion (compression) protocol takes about \(310\) \(\mu\)s, corresponding to engine efficiency, \(\eta = 42 \pm 6\%\), which is remarkably close to the Otto limit, in this case, \(\eta_{\text{Otto}}=44\%\)~\cite{Peterson2019}.

The experimental joint work and heat distribution, $P(W,Q)$ for a quantum Otto cycle implemented via NMR were reported in Ref.~\cite{Denzler2021}, where a detailed fluctuation theorem for a quantum cycle was also verified. In Ref.~\cite{Almeida2019b}, the authors investigate a spin quantum Otto cycle employing negative temperature states in NMR.

This specific proof-of-principle, implementation of an NMR quantum Otto engine, discussed in this section, does not exhibit any quantum advantages, although it can operate very close to the Otto efficiency, which is particularly hard to achieve in its classical devices counterpart. Some quantum advantages in thermal protocols will be discussed in the next sections. It is interesting to note that coherence's role in quantum thermal protocols is subtle. In the model presented here (which involves complete thermalization), the coherence contributes to entropy production as part of the efficiency lag, as seen in Eq.~(\ref{eq:efficience_lag}). On the other hand, it has been theoretically observed that for a quantum Otto cycle with partial thermalizations, coherence can improve the engine performance throughout a kind of dynamical interference ~\cite{Camati2019}.".

\subsection{Quantum refrigerators with indefinite causal order }

Cooling quantum systems is of significant importance to many applications in quantum technologies, and several methods to perform such a thermodynamic task on quantum devices in fast and efficient ways have been proposed in the literature. One particular method for refrigeration that showed new counter-intuitive predictions is related to the quantum-controlled switch of the application order of two or more quantum maps \cite{felce2020quantum,felce2021refrigeration,rubino2021quantum,nie2022experimental,dieguez2022thermal}. 

A quantum process is considered to have an indefinite causal order (ICO) if it is impossible to locally describe it as a convex mixture of the operations performed with definite causal orders \cite{oreshkov2012quantum,chiribella2013quantum,araujo2015witnessing}. 
Processes with ICO have been experimentally explored in contexts such as quantum computation \cite{ procopio2015experimental} and communication~\cite{ wei2019experimental, guo2020experimental,rubino2021experimental}. For more ICO experimental realizations, we refer the reader to a recent review in Ref.~\cite{goswami2020experiments}. 

In a recent investigation of a possible indefinite thermodynamic arrow of time~\cite{rubino2021quantum}, the effective quantum-controlled superposition of a heat engine and power-driven refrigeration was employed to demonstrate how quantum interference effects can be used to reduce undesired thermal fluctuations~\cite{rubino2021quantum}. In Ref.~\cite{guha2020thermodynamic}, it was demonstrated that the application of two different thermal channels in causally inseparable order can enhance the potential to extract work when compared to their definite order of occurrence. For a given working substance in a quantum state, described by the density operator $\rho$, work can be extracted via a cyclic process coupling it to and decoupling it from an external source. The cyclic dynamic of the system is governed then by a time-dependent Hamiltonian $H(t)$ in the interval $t_i\leq t\leq t_f$ such that $H(t_i)=H(t_f)=H$. The maximal work that can be extracted in such a scenario is quantified by the {\it ergotropy} defined as
\begin{equation}
W(\rho)=\max_{U\in \mathcal{U}}\text{tr}[H(\rho-U\rho U^{\dagger})],    
\end{equation} 
where $U$ means the set of unitary transformations generated by this particular time-dependent Hamiltonian~\cite{simonov2022work}. In this way, by electing the ergotropy as a figure of merit, the authors in Ref.~\cite{simonov2022work} showed that the activation of quantum maps through the quantum
switch always entails a non-negative gain in ergotropy when compared to their consecutive application. In particular, it was demonstrated that a nonzero work can be extracted from a
system thermalized by two coherently controlled reservoirs \cite{simonov2022work}. Thermal devices powered by generalized measurements with ICO were also presented in Ref.~\cite{dieguez2022thermal} where nontrivial effects coming from the interference of causal orders were explored to fuel a heat engine, thermal accelerator, as well as in a refrigerator cycle. 

In the following, we describe the experiment reported in Ref.~\cite{nie2022experimental} consisting of an NMR implementation of the power-driven refrigerator based on indefinite causal order. This interesting possibility reveals how the intrinsic indefiniteness in the causal order structure revealed by a quantum-controlled process can be used to perform a task that could not be realized using the same operations with a definite causal order~\cite{felce2020quantum,felce2021refrigeration}, and then which suggests a new kind
of non-classical resource for a thermodynamic task~\cite{felce2020quantum,rubino2021quantum} that can be experimentally investigated.

Different orders of applications for two consecutive thermalizations with reservoirs in the same temperature are completely redundant, due to the fact that any definite order will lead to the same Gibbs distribution as an output state. However, this is not the case if both operations are taken with an ICO. As we discuss in the following, non-classical effects coming from the interference of such ICO can be explored to perform a refrigerator cycle.
Then, employing the switch of two thermalizing maps, the authors of Ref.~\cite{nie2022experimental} constructed a single cycle of the ICO refrigerator, and evaluate its efficiency by measuring the work consumption and the heat extracted from a low-temperature (cold) reservoir Ref.~\cite{nie2022experimental}.

Theoretically, it is possible to effectively represent a thermalization process as the action of two depolarizing channels $\mathcal{T}^{1}$ and $\mathcal{T}^{2}$, each one with their respective Kraus operators $\{E^1_i\}$ and $\{E^2_j\}$.
In this setting, one's can construct the Kraus decomposition of the resulting map from the quantum switch of channels $\mathcal{T}^{1}$ and $\mathcal{T}^{2}$ as described by the following operators acting on the composite state of the working substance and the ancilla (the order control system) as~\cite{nie2022experimental}
\begin{equation}\label{Eq:decmp}
    K_{ij}:=E^2_jE^1_i\otimes\ket{0}\bra{0}+E^1_iE^2_j\otimes\ket{1}\bra{1},
\end{equation}
where $E^1_i$ and $E^2_j$ denote the Kraus operators acting on the working system for the formalization channels $\mathcal{T}^{1}$ and $\mathcal{T}^{2}$, respectively, and the projectors $\ket{\ell}\bra{\ell}$ ($\ell=0,1$) lie in the quantum controller space. Surprisingly, it was theoretically demonstrated in ~\cite{felce2020quantum} that even if both thermalization processes occur with the same temperature, the output state after applying the switch can vary, depending on the projective measurement result on the order-control ancilla if we start the experiment with an ancilla with some degree of coherence in the computational basis. By initializing then the ancilla with maximum coherence in the computational basis, this is, starting with $\rho_{c}=\ket{c}\bra{c}$ such that $\ket{c}=\frac{1}{\sqrt{2}}(\ket{0}+\ket{1})$, the output state can be calculated as
\begin{equation}
     \rho^{\text{sw}}=\mathcal{E}(\mathcal{T}^{1},\mathcal{T}^{2})(\rho_w\otimes\rho_{c})=\sum_{ij}K_{ij}(\rho_w\otimes\rho_{c})K_{ij}^{\dagger},
\end{equation}
with $\rho_w$ being the initial state of the working substance in the mentioned cycle and $\mathcal{E}(\mathcal{T}^{1},\mathcal{T}^{2})$ is the order switch map described by the Kraus operator in Eq.~(\ref{Eq:decmp})~\cite{nie2022experimental}. If we consider both thermalizations with the same temperature $\mathcal{T}^{1}=\mathcal{T}^{2}=\mathcal{T}$, one can show that the final state after the quantum switch is
\begin{equation}
    \rho^{\text{sw}}= \frac{1}{2}[\rho_{T}\otimes\mathbb{I}+\rho_{T}\rho\rho_{T}\otimes(\ket{0}\bra{1}+\ket{1}\bra{0})],
\end{equation}
where $\rho_{T}=e^{-H/k_bT}$ is the thermal equilibrium state at temperature $T$ with a two-level system Hamiltonian $H=\delta\ket{e}\bra{e}$ (the $\ket{e}$ representing the excited state). It is clear by looking at the above result that if the ancilla is projected in the computational basis, or if we just trace out the controller system, the working substance will be in a thermal state, this is, $\rho_T$. In both cases, projecting the ancilla means that we are choosing one of the definite orders, or taking the incoherent mixture of them (which are equivalent to the projection result, since we are dealing with the same thermalization). On the other hand, if we choose to project the ancilla in a complementary basis $\{\ket{+},\ket{-}\}$, we will be able to see an interference effect coming from the ICO. The output state then reads
\begin{equation}
    \rho_{\pm}=\frac{\rho_T\pm \rho_{T}\rho\rho_{T}}{2p_{\pm}},
\end{equation}
with $p_{\pm}=\text{tr}[(\rho_T\pm \rho_{T}\rho\rho_{T})/2]$ being the probabilities for each possible ancilla projection. After this projection on the control system, the switch can lead the working substance with an effective temperature that can be higher or lower than $T$. Even if the system starts with a temperature equal to the reservoirs, the final result can still be different due to the interference of the ICO.

 The refrigerator cycle and its corresponding NMR experimental investigation in Ref.~\cite{nie2022experimental} were designed as follows. A four-qubit system, including the working substance, the controller, and the two heat sources (that play the role of thermal reservoirs) starts in a state equivalent to $\rho_0=\rho_T\otimes \rho_c\otimes\rho_T\otimes\rho_T$. In the reported experiment, the initial effective
temperature of the working substance is settled to be the same as the reservoirs (prepared in equivalent Gibbs states). The switch operation, in order to implement the ICO, is decomposed by a
concatenation of rf-control pulses and free Hamiltonian evolution~\cite{nie2022experimental} employing a Toffoli gate. The experiment was implemented using four spin-1/2 nuclei of the molecule $^{13}C$-iodotrifluoroethylene ($\mathrm{C_{2}F_{3}I}$) dissolved in acetone $\mathrm{D}_{6}$ to realize the ICO refrigerator. The $^{13}C$ was employed as an ancilla to control the indefinite causal order procedure, while one of the $^{19}F$ acted as a working substance with the remaining two $^{19}F$ playing the role of the thermal reservoirs. The Experiment was performed at room temperature on a Bruker AVANCE $600$ MHz NMR spectrometer equipped with a cryogenic probe. The natural Hamiltonian of the system in their experiment can be written as
\begin{equation}
    H=-\sum_{i=1}^{4}\omega_i\frac{\sigma_z}{2}+\sum_{i<j}^{4}\pi J_{ij}\sigma_{i}^{z}\sigma_{j}^{z},
\end{equation}
with $\omega_i/2\pi$ being the Larmor frequency of the $i$-th spin, and $J_{ij}$ being the scalar coupling between the $i$-th and $j$-th spins~\cite{nie2022experimental}.

 They investigated the probability distribution of each measurement outcome of the quantum-controlled ancilla, which determines the success probability for a
particular heat flow direction. The quantum interference of the causal order is enhanced as the temperature goes lower and it is more likely to happen with the $\ket{+}$ projection, which is related to a cool down of the
working substance~\cite{nie2022experimental}. The amount of heat
exchanged between the working substance and reservoirs can be identified as
\begin{equation}
    \Delta Q^{ICO}_{\pm}=p_{\pm}\left( \text{tr}[(\rho_{\pm}-\rho_T)H]\right),
\end{equation}
where $\rho_{\pm}$ is the post-measured state for the working substance after the ancilla projection. The heat flow which provides
resources to construct quantum refrigerators is related to the $\ket{-}$ output, while the opposite result can be used to construct heat engines. The COP is then evaluated in terms of Maxwell's
demon mechanism, in which the work consumption and
heat extraction can be straightforwardly quantified \cite{nie2022experimental}. The ICO process is realized and the refrigerator cycle is viable when the desirable measurement result occurs, otherwise, the cycle is reinitialized. Classical heat exchange occurs with the reservoirs. Finally, since the thermalizations are isochoric processes, the consumption of work happens with the initialization of the ancilla and the erasure of Maxwell's demon's memory. To evaluate the COP, it was measured the heat energy extracted from the cold reservoir to the heat bath, and the
work consumption in a given cycle~\cite{nie2022experimental}.

The reported experiment showed a genuinely distinct behavior to what is known as the COP of a classical Carnot refrigerator, which approaches infinity when the limit of the cold and hot reservoirs approaches the same temperature. The COP of the ICO refrigerator reaches an optimal finite value in that limit, revealing, then, the peculiarities that one can expect to have when dealing with ICO-based cycles. With this extreme limit of investigation, their experiment demonstrates then that the ICO process may offer a new resource with non-classical heat exchange, and paves the way towards the construction of quantum refrigerators based on non-classical features~\cite{nie2022experimental} verified with the employment of NMR techniques.

\subsection{Quantum engines powered by measurements }

Measurements acting upon a quantum system do not generally preserve the measured system as it happens in classical physics, then it is possible, for example, to induce a change in the internal energy of some quantum working substance by making measurements on it. This possibility has been recently explored to develop a new class of thermal devices, known as thermal devices powered by quantum measurements~\cite{Talkner2017,Elouard2017,Brandner2015, Lin2021}. 
For instance, employing non-selective measurements, it was proposed a single temperature heat engine without the need for feedback control to extract work~\cite{Talkner2017}.
This cycle can be conceived by replacing the standard hot thermal reservoir of an Otto cycle with a measurement device that is responsible for providing heat to the working substance. 

The concept of quantum heat was identified as the stochastic energy fluctuations taking place during a quantum measurement \cite{Elouard2017-1}. Moreover, several measurement-based thermodynamic protocols have been studied in the last few years~\cite{Bresque2021,Campisi2019,Ding2018,Jordan2020,Mohammady2017,Campisi2017,chand2017single,chand2017measurement,chand2018critical,anka2021measurement}.

More recently, it was realized the possibility to perform a heat engine cycle by replacing the projective measurements with a more general notion of measurements \cite{behzadi2020quantum}. Generalized measurements can include phenomena that can not be described using only projective measurements. A particularly important application to thermodynamics can be identified with the employment of weak measurements~\cite{jacobs09,alonso16,pati2020,mancino18} since generalized measurements can be thought of as a sequence of weak measurements~\cite{oreshkov05,Dieguez18,pan16}. 
Weakly measured systems are relevant for consistent observations of work and heat contributions to an externally driven quantum stochastic evolution~\cite{alonso16}. From the experimental point of view, work and heat dynamics along single quantum trajectories, as well as information dynamics of a quantum Maxwell's demon, were investigated by employing superconducting qubits weakly coupled to a meter system~\cite{naghiloo2020heat,naghiloo2018information}. 
Also, a new class of Maxwell's demon was realized with a tunable dissipative strength using nitrogen-vacancy center~\cite{hernandez2022autonomous}.

\begin{figure}[h]
\centering
	\includegraphics[scale=0.70]{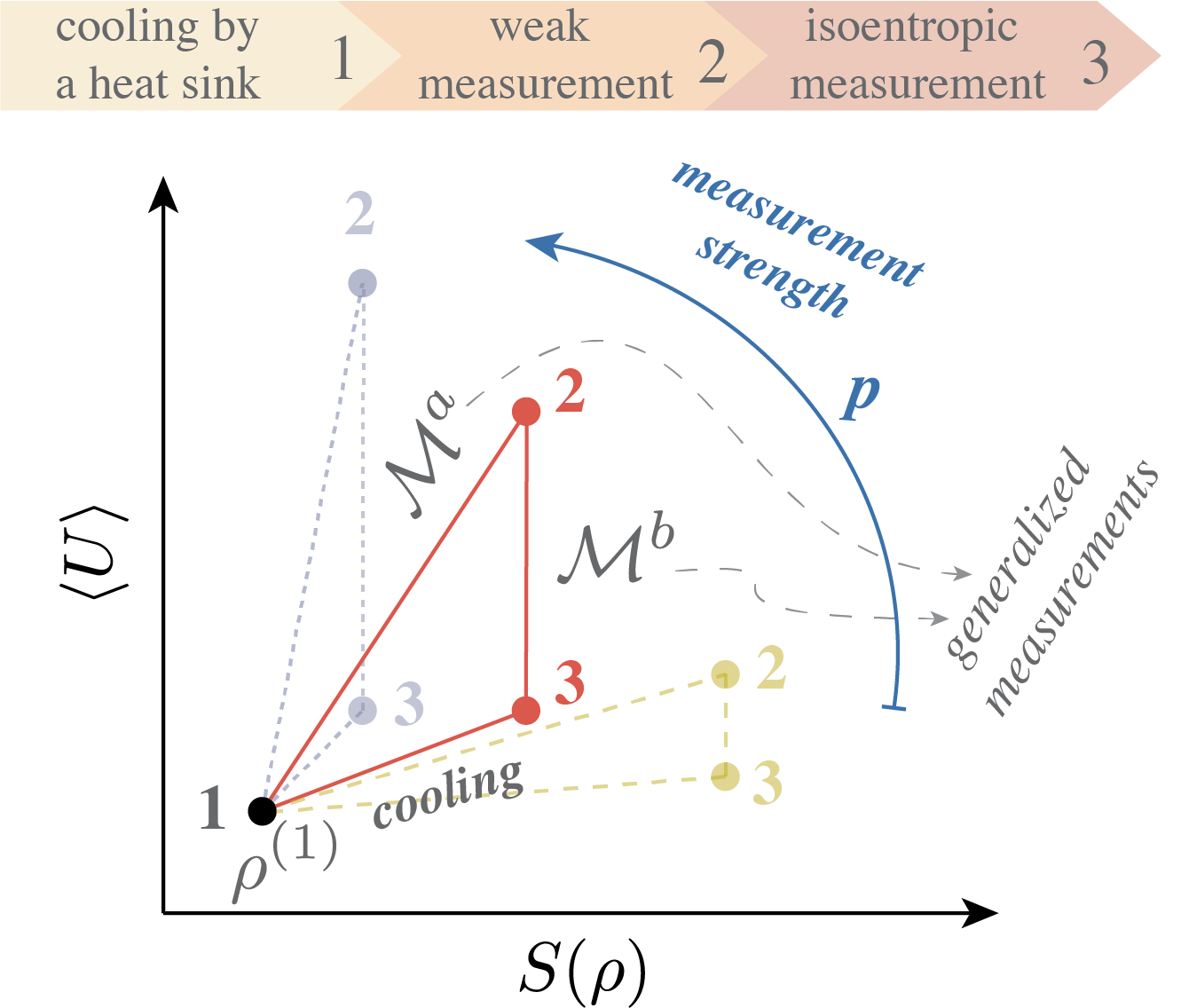}
	\caption{Schematic description of the heat engine cycle powered by generalized measurements. First, the work substance interacts with a cold sink and it is cooled to the equilibrium state $\rho^{(1)}$. Second, a general measurement channel $\mathcal{M}^{a}$ is performed which leads to an increase in the average energy and von Neumann entropy of the working substance. Over this channel, the working substance absorbs energy in a stochastic way (heat) from the meter. Third, another general measurement channel $\mathcal{M}^{b}$ chosen properly to produce an isentropically decreasing of the average energy of the working substance, is applied. The energy exchanged with the meter in the later channel has a work nature. The scheme also illustrates the variation of the internal energy and entropy for different measurement strengths.Adapted from Ref.~\cite{lisboa2022experimental}.}
	\label{maqmed}
\end{figure}

In the following, we describe a proof-of-concept experiment of a spin heat engine powered by generalized measurements that were performed employing NMR platform~\cite{lisboa2022experimental}. The cycle consists of two non-selective generalized measurements channels with adjustable measurement strengths, one dedicated to fueling the device, then playing the role of a heat source, and the other committed to work extraction when applied in an isentropic way~\cite{behzadi2020quantum} as depicted in Fig.~\ref{maqmed}. By changing the internal energy in an isentropic way, the working substance is considered informationally closed in a way that this operation can be recognized as work extraction \cite{lisboa2022experimental}.

In the experiment reported in Ref.~\cite{lisboa2022experimental}, the authors used a liquid sample of $^{13}$C-labeled Sodium formate (HCO$_2$Na) diluted in deuterium oxide (D$_2$O). To look over a generalized measurement-powered cycle, the experimental control was focused on the $^{1}$H and $^{13}$C nuclei which have nuclear spins $1/2$. This sample was prepared with a low dilution ratio ($\approx 2\%$), such that the inter-molecular interactions can be neglected and the sample can be considered as a set of almost identically prepared pairs of two spin-$1/2$ systems. The $^{18}$O has nuclear spin $0$, $^{23}$Na has nuclear spin $3/2$, and both nuclei played no important role in their experiment.

The $^{13}$C nuclear spins were initially cooled by performing spatial average techniques \cite{serra2014,Micadei2019,Micadei2021},
resulting in a state equivalent to the Gibbs state $\rho^{(1)}=\text{exp}\left[-\beta H^{\text{C}}\right]/\mathcal{Z}$,
at cold inverse spin temperature $\beta=1/\left(k_{B}T\right)$, where
$\mathcal{Z}$ is the partition function. The system's Hamiltonian is associated with an offset of the transverse rf fields with relation to the resonance of $^{13}$C nucleus, given by $H^{\text{C}}=-\left(h\nu/2\right)\sigma_{z}$ \cite{Micadei2021}, with $\sigma_{\ell}$, $\ell=\left(x,y,z\right)$
standing for the Pauli matrices wherein they adjusted the frequency as $\nu \approx 1$~kHz.

In the second stroke of this cycle, the authors performed a generalized non-selective measurement on the working
substance ($^{13}$C nucleus), which is effectively implemented by a controlled interaction with an ancillary system (the proton), prepared in an initial state equivalent to $\rho^{\text{H},0}=\ket 0 \bra 0$. The ancillary system is not observed and after the interaction, it can be traced out. The interaction with the ancillary system is managed to assemble the desired measurement map, leading to, 
$\mathcal{M}^{a}:\rho^{(1)} \rightarrow \rho^{(2)}=\sum_{i}{\mathcal{M}_{i}^{a}}\rho^{(1)}{\mathcal{M}_{i}^{a}}^{\dagger}$
where the Kraus operators of the generalized measurement are ${\mathcal{M}^{a}_{1}}=\sqrt{1-p\Omega}|0\rangle\langle0|+|1\rangle\langle1|$
and ${\mathcal{M}^{a}_{2}}=\sqrt{p\Omega}|1\rangle\langle0|$ (satisfying
$\sum_{i}{\mathcal{M}_{i}^{a}}^{\dagger}{\mathcal{M}_{i}^{a}}=I$), and $\Omega=1-e^{-\beta h\nu}$
is a factor that depends on the system temperature after the first stroke. The parameter $p$ is related to the measurement strength. 

In their implementation, the range 
$0<p\Omega<1$ varies from a weak to a moderated measurement strength. In this stage, the internal energy variation of the working substance can be interpreted as the heat absorbed from the meter, whenever its von Neumann entropy increases, as usually adopted~\cite{behzadi2020quantum} in this measurement-based cycle. So, the heat absorbed in stroke two is simply given by 
$\langle \mathcal{Q}^{p}\rangle=\text{tr}\left[H^{\text{C}}\left(\rho^{(2)}-\rho^{(1)}\right)\right]$, being directly associated to the variation of the nuclear spin magnetization. 
A second non-selective generalized measurement channel was performed, in the third stroke, on the working substance leading to $\mathcal{M}^{b}:\rho^{(2)}\rightarrow\rho^{(3)}=\sum_{j=1}^{2}\mathcal{M}^{b}_{j}\rho^{(2)}{\mathcal{M}^{b}_{j}}^{\dagger}$
with the Kraus operators $\mathcal{M}^{b}_{1}=|0\rangle\langle0|+\sqrt{1-q}|1\rangle\langle1|$
and $\mathcal{M}^{b}_{2}=\sqrt{q}|0\rangle\langle1|$.

By imposing 
\begin{equation}
 q=\frac{\left(2p-1\right)\Omega}{\left(p-1\right)\Omega+1},
 \label{parameter}
\end{equation}
the authors ensured that the von Neumann entropy of the working substance does not change during the measurement channel $\mathcal{M}^{b}$, $\Delta S^{b} = \text{S}(\rho^{(3)})-\text{S}(\rho^{(2)})=0$. In this way, the energy exchanged in stroke three can be associated to
work delivered by the working substance to the meter \cite{behzadi2020quantum}.
The work performed over the working substance by the measurement channel $\mathcal{M}^{b}$ is then given by 
$\langle \mathcal{W}\rangle=\text{tr}\left[H^{\text{C}}\left(\rho^{(3)}-\rho^{(2)}\right)\right]$. 

\begin{figure}[ht]
\centering
\includegraphics[scale=0.40]{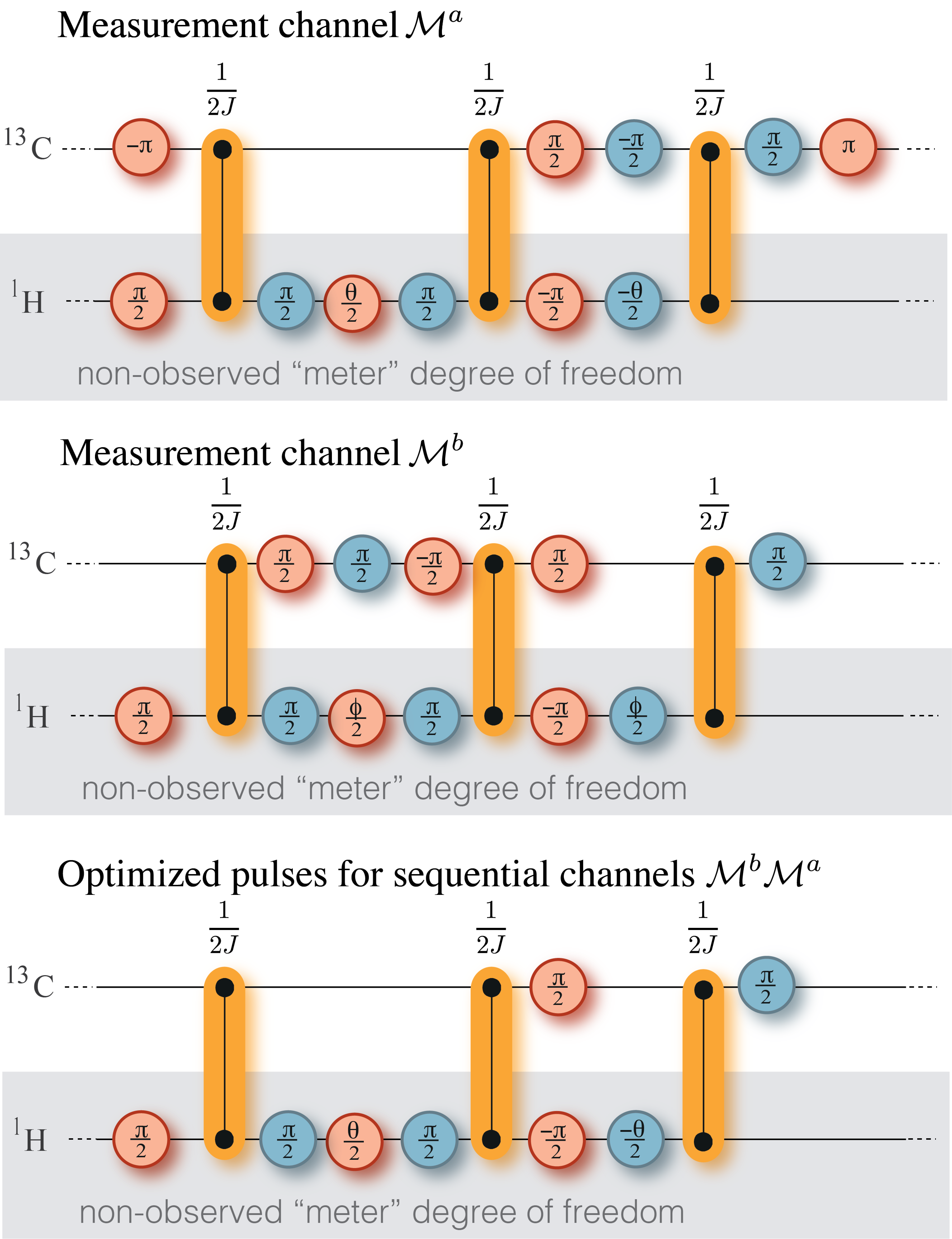}
\caption{Pulse sequences for the generalized measurement powered engine. The non-selective generalized measurement channel $\mathcal{M}^a$ (second stroke) is effectively performed using the sketched pulse sequence. The measurement strength parameter, $p$, is tuned by the angle $\theta \in [0,\pi/2)$ according to $\theta=\arccos{(1-2 p \Omega )}$. The channel $\mathcal{M}^b$ (third stroke) can be effectively implemented throughout the sketched pulse sequence. The strength parameter $q$ is tuned by the angle $\phi \in [0,\pi/2)$ using the relation $\phi=\arccos{(1-2q)}$. The pulse sequence for the consecutive application of $\mathcal{M}^a$ and $\mathcal{M}^b$ is obtained by setting the strength $q$ as a function of $p$. It was optimized, using the commutation relations for Pauli rotations to cancel redundancies, resulting in a shortened effective sequence that also minimizes errors. After the second and third strokes, the magnetization of the $^{13}$C nuclear spin is measured (determining the internal energy variation) and a QST is performed. From the last data, it was obtained the variation of the von Neumann entropy. The $^{1}$H nuclear spin was not observed resulting in the desired non-selective generalized measurement channels (the CPTP maps) acting on the $^{13}$C nucleus. Adapted from Ref.~\cite{lisboa2022experimental}.}	\label{fig:pulse_measure}
\end{figure}

The experiment \cite{lisboa2022experimental} was performed using a Varian 500 MHz spectrometer equipped with a double-resonance probe head with a magnetic field-gradient coil. The states of the nuclear spins were controlled using time-modulated rf pulses in the transverse direction, longitudinal magnetic field gradients, as well by sequences of free evolution of the system under the action of the scalar coupling. The latter is described by the Hamiltonian $H_J=({\pi\hbar}J/4)\sigma_z^{\text{H}}\otimes \sigma_z^{\text{C}}$, where $J \approx 194.65$~Hz is the coupling constant between the $^{1}$H and $^{13}$C. The generalized measurement channels $\mathcal{M}^a$ and $\mathcal{M}^b$ that act on the $^{13}$C nuclear spin were implemented through the interaction with the $^{1}$H nuclear spin which played the role of the internal degree of the meter, as the pulse sequence sketched in Fig.~\ref{fig:pulse_measure}. At the end of the experiment, the state of the $^{1}$H nuclear spin was not observed \cite{lisboa2022experimental}. So, after the pulse sequence, it was only observed the averaged effects of the interaction with the $^{1}$H on the $^{13}$C dynamics, leading to the desired non-selective generalized measurement map acting on the working substance ($^{13}$C) \cite{lisboa2022experimental}.

\begin{figure}[h]
\centering
\includegraphics[scale=0.40]{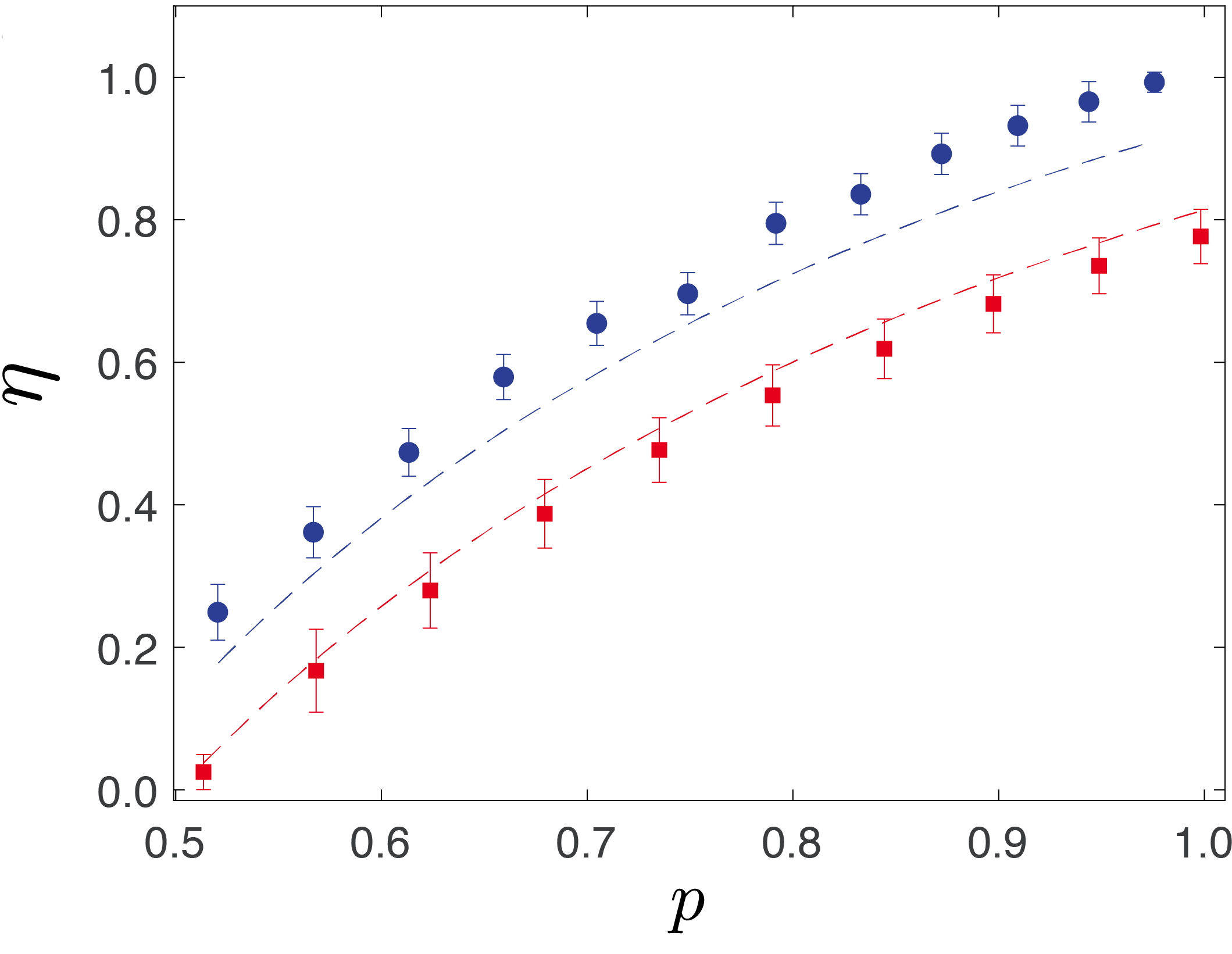}
\caption{Efficiency of the generalized measurement-powered heat engine as a function measurement strength $p$. The dashed curves represent numerical simulations of the cycle including some non-idealities. The symbols are the observed experimental data with the red squares associated with the initial spin temperature $k_BT_1 = 1.88\pm 0.21$~peV while the blue circles correspond to the initial spin temperature $k_BT_2 = 2.98+\pm 0.19$~peV. Adapted from Ref.~\cite{lisboa2022experimental}.}
	\label{effmed}
\end{figure}

An interesting point which was highlighted in the experimental results reported in Ref.~\cite{lisboa2022experimental}, is that this kind of quantum thermal device can reach unit efficiency while also achieving maximum extracted power at the same time with the fine-tuning of the measurement strengths, as shown in Fig.~\ref{effmed}. A near to unity efficiency is obtained when $p\rightarrow 1$, which corresponds to the strength of the measurement achieving the value $\Omega$ which is related to the excited population of the initial Gibbs distribution. 

It is important to note that, due to the intrinsically quantum nature of the measurement back-action, it is not possible to establish a direct comparison with classical thermal cycles without feedback control wherein the traditional thermodynamic roles are expected. Both the hot heat source and the work extraction procedure here are obtained by using a quantum generalized measurement which leads the system's state to non-equilibrium after the interaction with the meter. Then, for instance, the resulting state after the measurement which plays the role of a heat source can be equivalent to a Gibbs state with effective negative temperature. All these genuine quantum effects related to the measurement back-action are explored to reach efficiency near unity as shown in Fig.~\ref{effmed}.

From a theoretical point of view, the efficiency of the measurement-powered cycle only depends on the measurement strength of the first measurement channel. Otherwise, in a practical scenario, precise knowledge of the initial state temperature and fine control of the measurement channel parameters are required to reach the theoretical prediction and the unit efficiency. In the implementation discussed in this section, due to experimental non-idealities and some uncertainty about the initial spin temperature, we observed that the efficiency also depends on the initial state (as in Fig.~\ref{effmed}). The efficiency reaches a value near the unit for one of the initial spin temperatures considered, showing that having sufficient control over the parameters involved, it is possible to implement the measurement-powered cycle with maximum efficiency. The simulated curves, on the other hand, consider the initial experimental state and simulate some of the expected errors in the pulse sequence implementation of each generalized measurement channel. 
The observed experimental data which behaved better than the numerically simulated curves should be interpreted as a particular situation where the experimental error contributes to obtaining a result slightly better than the simulation.

This NMR investigation could unveil practical applications of measurement-based protocols in quantum thermodynamics, as well as open new possibilities for efficiency-enhancing from the combination of a system driven by a time-dependent Hamiltonian and measurements in quantum devices \cite{lisboa2022experimental}.
 
\subsection{Quantum batteries}

In the previous sections, we have seen how to extract work from cyclic processes using QHEs. However, we may ask ourselves how we could store efficiently this extracted energy in another quantum system. That is exactly what are quantum batteries, energy-storing devices that exploit quantum features, like quantum superposition or entanglement, in order to speed up the process of charging (energy deposition) or discharging (energy extraction) of the battery \cite{Bhattacharjee2021, Campaioli2013, Alicki2013, Hovhannisyan2013}.

The simplest form of a quantum battery consists of a two-level quantum system \cite{Santos2021, Joshi2021}. If the system is in the ground state, we say that the battery is discharged, if it is in the excited state, we say that the battery is charged. Let us take a spin as the two-level quantum battery, we could charge it by applying an external magnetic field \cite{Binder2015} or coupling it with an ancillary spin that acts as a charger \cite{Le2018}. 

When using ancillary systems to charge a battery with \(N\) cells (e.g. each spin being a battery cell in the spin chain), we emphasize here two possible procedures, parallel and collective charging. In the parallel charging protocol, each battery cell is charged simultaneously under the action of one of the \(N\) chargers. On the other hand, the collective charging protocol considers all cells together as a battery pack, and all the chargers act simultaneously on it. Since it uses non-local `entangling' operations, collective charging protocols exhibit a quantum advantage over parallel ones that scale extensively \cite{Bhattacharjee2021, Binder2015}. However, it is important to mention that albeit the necessity of these non-local operations, there is no necessity for the battery cells to be entangled. 

\begin{figure}[h]
\centering
	\includegraphics[scale=0.55]{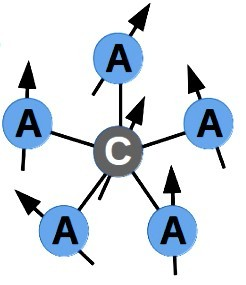}
	\caption{Star-topology network of spins, where we have a central spin (C) interacting individually with each ancillary spin (A). Figure adapted from the Ref.~\cite{Mahesh2021}.}
	\label{STR}
\end{figure}

In order to experimentally realize such quantum batteries in an NMR spectrometer, we have a well-known problem of scalability of the technique in quantum information (computation) applications, when considering a large number of battery cells or chargers, to face. One way to avoid this spectral complexity of using a large number of spins in NMR spectroscopy is to make use of the symmetries of the system. For instance, quantum registers are systems with multiple qubits and can be categorized by their network topology. A recent review by Mahesh et al. \cite{Mahesh2021} showed the advantages of using a particular type of quantum register: the star-topology register (STR) with spins qubits, see Fig.~\ref{STR}. 

An STR of \(N\) qubits consists of \(N-1\) ancillary qubits surrounding a single central qubit. In general, the STRs realized in liquid-state NMR experiments do not have any interaction among the ancillary qubits, which are indistinguishable from one another, and the central qubit is made of a different nuclear isotope than the ancillae. 

\begin{figure}[h]
\centering
	\includegraphics[scale=0.6]{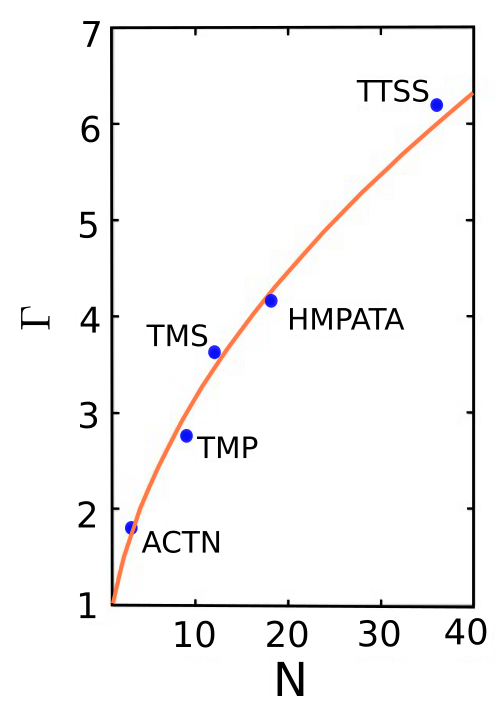}
	\caption{Quantum advantage \(\Gamma\) versus number of chargers \(N\). We see that both experimental data (dots) and theoretical curve (solid line) agree. Figure adapted from the Ref.~\cite{Joshi2021}.}
	\label{QBattery}
\end{figure}

These STRs were recently modeled as quantum batteries in a 500 MHz Bruker NMR spectrometer \cite{Joshi2021}. In the reported experiment, the authors used different samples (ACTN, TMP, TMS, HMPATA, and TTSS), with an increasing number of charger spins (\(N = 3, 9, 12, 18\) and \(36\), respectively). The collective charging scheme was investigated and the quantum advantage over the parallel scheme was reported. The quantum advantage \(\Gamma\) of the collective scheme is defined as the ratio $P_N/P_1$, where \(P_1\) and \(P_N\) are the charging powers of the parallel and collective charging schemes, respectively. Figure~\ref{QBattery} illustrates the quantum advantage \(\Gamma\) against the number \(N\) of chargers, where dots represent experimental data for the different systems used, and the solid line is the theoretically expected curve \(\Gamma = \sqrt{N}\). Despite both charging schemes being quantum mechanical by definition, the collective charging scheme is superior probably because it exploits quantum correlations among the spins. This fact is explored explicitly in Ref.~\cite{Joshi2021} by the realization of numerical calculations of the quantum discord and entanglement entropy during the processes, where at the end of the charging protocol, the spins get uncorrelated ~\cite{Joshi2021}. These findings suggest a relation between non-classical correlations and the advantage of the collective charging scheme. Notwithstanding, by introducing a load spin, it was shown the establishment of a \textit{quantum charger-battery-load} (QCBL) circuit, where the battery spin stored energy for up to two minutes, and yet it was able to transfer the stored energy to the load spin.

\subsection{Quantum thermometry }

Quantum thermometry, which is a subarea of quantum metrology, is associated with the precise measurements of temperature employing non-classical features. Determining the temperature of an object may not seem to be, at first sight, a challenging issue, however, doing it precisely in a scenario of very low values could be a non-trivial task even in the classical realm. In classical thermodynamics, the zeroth law of thermodynamics provides a way of tagging a value of temperature for two systems in thermal equilibrium with a fixed reference, while the second law of thermodynamics (throughout the reversibility of the Carnot engine) establishes a protocol to measure temperature ratios, thus removing any ambiguity left by the arbitrariness of the tag value established by the zeroth law \cite{Mehboudi2019}. Accordingly, standard thermometers infer the temperature of a system by letting the thermometer go to equilibrium with the system of interest and measuring another physical property in the thermometer.

When considering quantum thermodynamics, things got a little complicated as the measurement becomes more and more invasive and carries a measurement back-action. From the system's energy spectrum, we can estimate the temperature of the quantum system by performing global measurements provided that our quantum system is in thermal equilibrium and follows a Boltzmann distribution~\cite{Stace2010}. A less invasive way for realizing this measurement would be introducing a quantum probe (a third part system or degree of freedom) to interact with the system of interest. This probe would couple to the system through a very weak dissipative interaction until it reaches a thermalized state at the same temperature \(T\) as the system \cite{Correa2015}. This process, \textit{in principle}, does not significantly disturb the system, and both probe and system would end up being uncorrelated, this way we could measure any temperature-dependent property of the probe alone and fully estimate the system's temperature.

Nevertheless, none of the aforementioned methods brings up any quantum advantage in the measurement precision if compared to an ordinary (classical) thermometer \cite{Stace2010}. Quantum thermometry deals with the precision limit at which the temperature of a quantum system can be determined, and a true \textit{quantum thermometer} works in a distinct manner: it encodes the information about the system's temperature in a quantum property of the probe such as the relative phase, which can be estimated using interferometric methods. If the decoherence time of the probe is greater than the interaction time, we say we have a nonthermalizing thermometer \cite{Raitz2014}. 

Following into the NMR applications, Raitz et al.~\cite{Raitz2014} used a quantum scattering circuit with two qubits (a probe and a target qubit) to implement an NMR quantum thermometer. Nonetheless, these circuits are, in general, characterized by a probe qubit and a mixed state \(\rho\), where the probe qubit is initially prepared in the pure state (pseudo-pure state) \ket{0}, and a first Hadamard gate is applied to put the probe in the superposition $(\ket{0} + \ket{1})/\sqrt{2}$. Then a controlled unitary operator is applied to the system if the probe is in the state \ket{1}, a control-$U$ gate. Finally, another Hadamard gate is applied to the probe. This interferometric protocol encodes information about the system into the superposed probe state which is subsequently measured.

\begin{figure}[h]
\centering
	\includegraphics[scale=0.6]{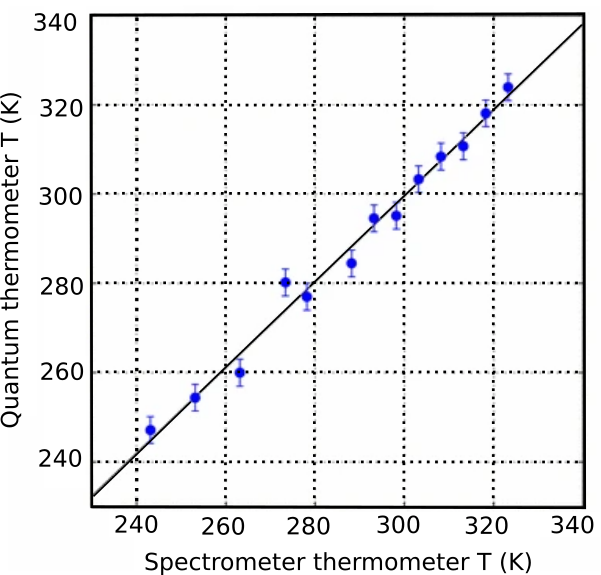}
	\caption{Temperature of the Hydrogen's nuclear spin measured with the quantum thermometer against temperature measured by the spectrometer. Figure adapted from the Ref.~\cite{Raitz2014}.}
	\label{QThermometer}
\end{figure}

In the Raitz et al.~\cite{Raitz2014} experiment a $^{13}C$-labeled chloroform molecule diluted in deuterated acetone, where the $^{13}C$ nuclear spin was used as the probe qubit, and the $^{1}H$ nuclear spin was the system with temperature to be measured. After pseudo-pure state preparation and calibration procedures, the quantum circuit was run for 13 different values of initial temperatures, which were measured using the quantum thermometer and also in the standard spectrometric form. The results are shown in Fig.~\ref{QThermometer}, where we observe a good agreement between the two methods.

The main advantage of using this quantum scattering circuits technique is that, independently of the size of the target quantum system, we can always encode the information about its temperature in a single probe qubit, and it is an alternative way of measuring temperature without the necessity of thermal contact.

Another implementation of a quantum thermometer in NMR platforms was also reported more recently in Ref.~\cite{Uhlig2019}. In this later paper, the authors used highly correlated quantum NOON states of spins to model a proof-of-principle quantum thermometer. The errors behind the measurements of temperature provided by the implemented thermometer, scaled approximately in the Heisenberg limit \(1/N\), which is more accurate than the standard quantum limit \(1/\sqrt{N}\), where \(N\) is the number of independent probes measured. To be more precise, it was shown an error scaling of \(N^{-0.94}\)~\cite{Uhlig2019}. This is the first implementation of a quantum thermometer working in the Heisenberg limit. Aside from precision improvements, some quantum thermometry strategies do not require reaching equilibrium between the temperature probe and the system.

\section{Information and thermodynamics} \label{Information and thermodynamics}

\subsection{Quantum Maxwell's demon }
Maxwell was one of the pioneers in addressing the consequences of statistical theory for thermodynamic rules. In 1867, he devised a thought experiment in which the second law would be apparently violated \cite{Maxwell_paper}. In view of testing this, Maxwell imagined a clever being capable of controlling the degrees of freedom of a macroscopic gas. Curiously, this intelligent being was later dubbed demon by Thompson \cite{Maxwell_book}.

\begin{figure}[h]
\centering
	\includegraphics[scale=0.4]{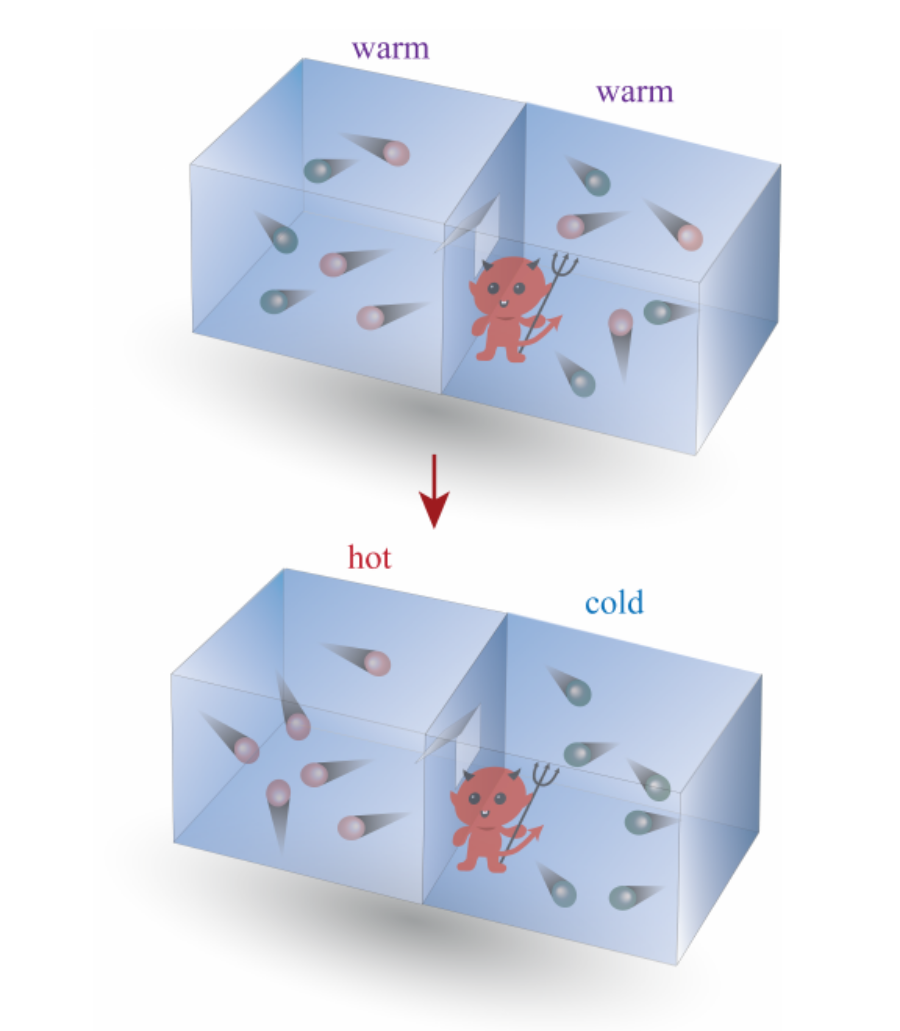}
	\caption{Maxwell's demon. The demon sorts the faster molecules to the left and the slow molecules to the right by opening and closing the gate in the middle at the right time. By this procedure, the demon is able to create and further increase a temperature gradient from a gas initially in thermal equilibrium. A macroscopic observer would see a spontaneous heat flow from the cold side to the hot side without any work performed, hence configuring an apparent violation of the second law of thermodynamics.}
	\label{MD}
\end{figure}

In this thought experiment, Maxwell supposed a gas in thermal equilibrium at temperature $T$ inside a box with adiabatic walls and two segregated parts, except by one gate between them as illustrated in Fig.~\ref{MD}. Supposing that an intelligent being or Maxwell's demon standing at the gate, like a doorman, able to open and close it at will, has the ability to monitor the velocity of each gas molecule and control a molecular flow from one side to another, see Fig.~\ref{MD}. 
Since the average kinetic energy of the particles is proportional to the temperature, at the end of this process, this means that the temperature of one side increases, and conversely the temperature of the other decreases. If this thought experiment were possible, a spontaneous heat transfer from a cold gas to a hot gas, without an external agent performing work, would be observed, apparently contrasting the second law of thermodynamics.

A fundamental aspect that Maxwell overlooked in his thought experiment was how exactly the demon obtains information about the velocities of the molecules. In 1929, Le� Szilard was one of the first who originally pointed out the inconsistency in this thought experiment \cite{Szilard,Szilard_2}, although his approach was not completely undoubted. Maxwell's demon must generate entropy when observing the velocities of molecules, storing and comparing velocity information, and opening and closing the door. If the demon's entropy is not taken into account, the system's entropy may appear to have reduced. However, given the entropy generated by the intelligent demon while executing the task, the global entropy is expected to grow. Szilard used a classical engine to address Maxwell's demon issue (see more in Sec. \ref{Szillard section}). A few years later, in 1951, Brillouin corroborated Szilard's assessment by demonstrating that a quantity of energy substantially bigger than $kT$ would be required if the demon employed photons to determine the position of the molecule~\cite{Brillouin}.

Maxwell's demon-\textit{gedanken} experiment produced a dilemma that persisted for almost 150 years~\cite{Maxwell_book}. In 1961, Rolf Landauer shed light on this conundrum. He proposed an erase principle to explain indubitably the increase in entropy~\cite{Landauer_1,Landauer_2}. According to Landauer's principle, memory erasure should be accompanied by entropy production, since it is an irreversible process. In 1982, Bennett utilized Landauer's idea to finally exorcise Maxwell's demon \cite{Bennett}. He realized that in order to complete the thermodynamic cycle, the demon's memory had to be erased, resulting in an entropy equal to the reduction associated with the passage from the cold to the hot body. Therefore, the second law of thermodynamics is restored by taking this entropy increase into account.

According to the preceding discussion, Maxwell's demon works as a kind of measurement-based feedback controller. It performs interaction with the system and thereafter performs one action depending on the measurement result. Recent analyses have considered the explicit change produced in the statistical description of the system as a result of the assessment of its microscopic information \cite{Parrondo}.
This has helped to understand that information-to-energy conversion can be governed by fluctuation theorems that hold on microscopic systems arbitrarily far from equilibrium~\cite{Camati2018b,Esposito,Jarzynski_2011,Sagawa_book},
besides enabling generalizations of the second law in the presence of a feedback controller~\cite{Sagawa_Ueda}.
Based on the trade-off between thermodynamics and information experimental ~\cite{Pekola2015,Camati_2016,Cottet_2017,Camati_QED,hernandez2022autonomous} and theoretical~\cite{Jarzynski_2012,Jarzynski_2014,Rupprecht_2019}
endeavors have been done in order to implement an efficient Maxwell\textquoteright s demon in the quantum framework. These investigations purposing to establish pragmatic applications within the current technological progress where information is an essential component to be manipulated.

\begin{figure}[th]
\centering
	\includegraphics[scale=0.4]{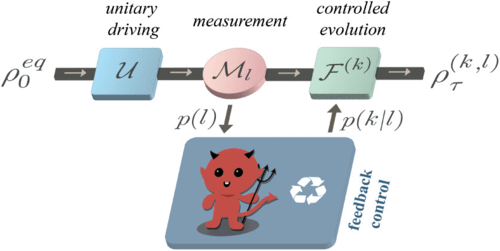}
	\caption{Maxwell's demon works as a measurement-based feedback controller. The system starts in the equilibrium state $\rho_{0}^{eq}$ and it is unitarily driven $U$ to a nonequilibrium state. Then the demon makes a projective measurement, $\mathcal{M}$, yielding the outcome $\ell$ with probability $p(\ell)$. The feedback operation $\mathcal{F}^{(k)}$ is applied with error probability $p(k|\ell)$. The environment temperature is kept fixed and the whole operation is much faster than the system decoherence time. Figure adapted from Ref.~\cite{Camati_2016}.}
	\label{MD_FB}
\end{figure}

One recent experiment in the setting of NMR spectroscopy used the trade-off between information theoretic quantities to implement an efficient Maxwell's demon in a quantum system~\cite{Camati_2016}. The authors experimentally assess the efficiency of this Maxwell's demon in rectifying entropy production owing to quantum fluctuations in nonequilibrium dynamics~\cite{Batalhao_2015}. Considering the scenario illustrated in Fig.~\ref{MD_FB}, the working system is a small quantum system, initially in the equilibrium (Gibbs) state $\rho_{0}^{eq}$ at inverse temperature $\beta=(k_{B}T)^{-1}$. Maxwell's demon is represented through a microscopic quantum memory. Suppose that the working system is driven away from equilibrium by a fast unitary time-dependent process, $U $, up to time $\tau_{1}$ (driving the system Hamiltonian from $H_{\tau_{0}}$ to $H_{\tau_{1}}$). The purpose of the control mechanism is to rectify the quantum fluctuations introduced by this nonequilibrium dynamics. To this end, the demon acquires information
about the system's s state through a complete projective
measurement, $\left\{ \mathcal{M}_{\ell}\right\} $, yielding the
outcome $\ell$ with probability $p(\ell)=\mathrm{tr\left[\mathcal{M}_{\ell}U\rho_{0}^{eq}U^{\dagger}\right]}$.
Based on the outcome of this measurement, a controlled evolution will
be applied. It will be described by unital quantum operations $\mathcal{F}^{(k)}$,
which may include a drive on the system's Hamiltonian
from $H_{\tau_{1}}$ to $H_{\tau_{2}}$. Furthermore, they
considered the possibility of error in the control mechanism, assuming
a conditional probability $p(k\vert\ell)$ of implementing the feedback
process $k$ (associated with the outcome $k$) when $\ell$ is the
actual observed measurement outcome. With a suitable choice of the operations $\left\{ \mathcal{F}^{(k)}\right\} $, the feedback control mechanism can balance out the entropy production due to the nonequilibrium drive
$U$. 

The main objective of this work was to perform a feedback-control
protocol in a quantum system and see the negativity of the mean entropy
production, experimentally witnessing Maxwell's demon in action.
The authors developed one mathematical equality for entropy production,
in the presence of feedback control, with experimental significance
for the successful design of Maxwell's demon, represented as
\begin{equation}
\langle\Sigma\rangle=-\mathcal{I}_{gain}+\langle {S}_{KL}\left(\rho_{\tau_{2}}^{(k,\ell)}||\rho_{\tau_{2}}^{(k,eq)}\right)\rangle+\langle\Delta {S}^{(k,\ell)}\rangle_{\mathcal{F}},\label{eq:4}
\end{equation}
which contains only information-theoretic quantities on the r.h.s.
The information gain $\mathcal{I}_{gain}=S(\rho_{\tau_{1}})-\sum_{\ell}p(\ell)S(\rho_{\tau_{1}}^{\ell})$
quantifies the average information that the demon obtains by reading
the outcomes of the measurement $\mathcal{M}$ \cite{Groenewold_1971,Lindblad_1972}, with $\rho_{\tau_{1}}=U\rho_{0}^{eq}U^{\dagger}$
being the system's state before the measurement; $\rho_{\tau_{1}}^{\ell}$ the
$\ell$-th post-measurement state which occurs with probability $p(\ell)$,
and $\mathcal{S}(\rho)$ the von Neumann entropy. The Kullback-Leibler (KL)
relative entropy, ${S}_{KL}\left(\rho_{\tau_{2}}^{(k,\ell)}||\rho_{\tau_{2}}^{(k,eq)}\right)=\mathrm{tr}\left[\rho_{\tau_{2}}^{(k,\ell)}\left(\ln\rho_{\tau_{2}}^{(k,\ell)}-\ln\rho_{\tau_{2}}^{(k,eq)}\right)\right]$,
expresses the information divergence between the resulting state of
the feedback-controlled process, $\rho_{\tau_{2}}^{(k,\ell)}$, and
the equilibrium state for the final Hamiltonian $\mathcal{H}_{\tau_{2}}^{(k)}$
in the $k$-th feddback process, $\rho_{\tau_{2}}^{(k,eq)}=e^{-\beta\mathcal{H}_{\tau_{2}}^{(k)}}/Z_{\tau_{2}}^{(k)}$.
The last term, $\langle\Delta {S}^{(k,\ell)}\rangle_{\mathcal{F}}=\langle {S}(\rho_{\tau_{2}}^{(k,\ell)})-{S}(\rho_{\tau_{1}}^{(\ell)})\rangle$,
represents the averaged change in von Neumann entropy due to the quantum
operation $\mathcal{F}^{(k)}$. The hallmark of Eq.~(\ref{eq:4})
is the possibility to characterize experimental entropy production
by only measuring information-theoretic quantities.

In the experimental implementation reported in Ref.~\cite{Camati_2016}, the authors assumed that the system and
the controller are two nuclear spins in Chloroform
molecules (CHCl3) in an NMR sample. The controller (or memory) qubit
is the spin-1/2 nucleus of the $\mathrm{^{1}H}$ and it plays
the role of Maxwell's demon. The Hydrogen effectively stores the measurement
outcomes on an orthogonal basis, which is then used to implement the
feedback control. On the other hand, the system is given by the spin-1/2
nucleus of the $\mathrm{^{13}C}$.

\begin{figure}[th]
\centering
	\includegraphics[scale=0.42]{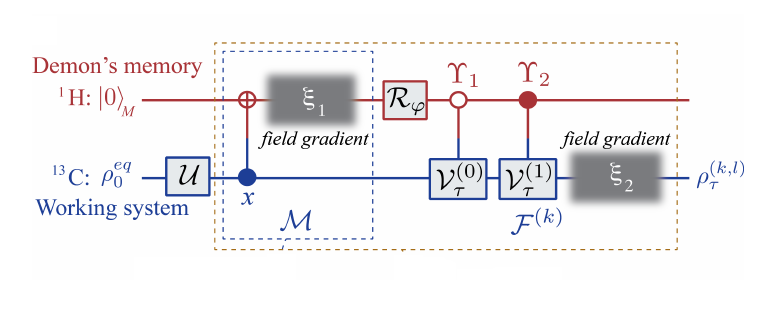}
	\caption{Quantum circuit for Maxwell's demon which was performed to measure all the information-theoretic quantities in rhs of the Eq.~(\ref{eq:4}). The controller (or memory) qubit is the spin-1/2 nucleus of the Hydrogen $\mathrm{^{1}H}$ and it plays the role of Maxwell's demon. The system is given by the spin-1/2 nucleus of the Carbon isotope $\mathrm{^{13}C}$. After the sudden quench $U$, information is acquired by the demon via an effective non-selective projective measurement in the energy basis of the $\mathrm{^{1}H}$. The control mismatch is implemented by a spin rotation $\mathcal{R}_{\phi}$, along the $x$ direction of the $\mathrm{^{1}H}$ nuclear spin, and the feedback operations are implemented by means of conditional evolutions $\Upsilon_{1}$ and $\Upsilon_{2}$. Figure adapted from Ref.~\cite{Camati_2016}.}
	\label{MD_FB_2}
\end{figure}

The feedback mechanism employed is sketched in Fig.~\ref{MD_FB_2}, where the whole feedback operation is much faster than the typical
decoherence times~\cite{Camati_2016,serra2014}. After the sudden quench $U$, information is acquired by the demon via the
natural $J$ coupling between $\mathrm{^{13}C}$ and $\mathrm{^{1}H}$
nuclei, under a free evolution, lasting for
about $6.97$ ms (equivalent to a CNOT gate). An effective non-selective
projective measurement in the energy basis of $H_{\tau_{1}}$
is accomplished with an additional longitudinal field gradient, $\xi_{1}$
(applied during $3$ ms). It introduces a full dephasing on the $z$
component of the memory state. This free evolution followed by dephasing
correlates the state of the working system ($\mathrm{^{13}C}$) with
the demon\textquoteright s memory ($\mathrm{^{1}H}$) leading to a
joint \textquotedblleft postmeasurement\textquotedblright{} state
equivalent to $|0\rangle_{\mathrm{H}}\langle0|\mathcal{M}_{0}\rho_{0}^{C}\mathcal{M}_{0}+|1\rangle_{\mathrm{H}}\langle1|\mathcal{M}_{1}\rho_{0}^{C}\mathcal{M}_{1}$,
where $\mathcal{M}_{0}$ and $\mathcal{M}_{1}$ are eigenbasis projectors
for $H_{\tau_{1}}$ with experimentally probed
outcome probabilities, $p(\ell)=50.0\pm0.4\%$ for $\ell=0,1$, as expected for the sudden quench implemented.

\begin{figure}[h]
\centering
	\includegraphics[scale=0.57]{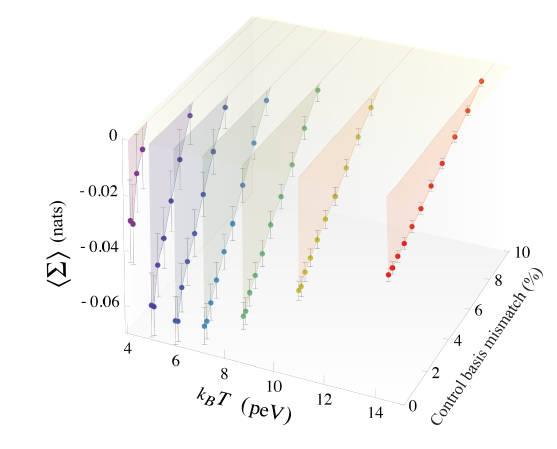}
	\caption{Experimental entropy rectification. Average nonequilibrium entropy production $\langle\Sigma\rangle$ in the measurement-based feedback protocol as a function of the initial temperature $k_{B}T$ and the control basis mismatch $p(k|\ell)$. The negative values are associated with entropy rectification by Maxwell's demon, according to Eq.~(\ref{eq:4}). Figure adapted from Ref.~\cite{Camati_2016}.}
	\label{MD_FB_3}
\end{figure}

Entropy rectification is achieved by a controlled evolution of the $\mathrm{^{13}C}$ nuclear spin guided by the demon's
memory (encoded in the $\mathrm{^{1}H}$ nuclear spin state). Such conditional evolution is implemented by means of the conditional evolutions $\Upsilon_{1}$ and $\Upsilon_{2}$, represented in Fig.~\ref{MD_FB_2}, and produce an
ideally controlled transformation $\Upsilon_{1}\Upsilon_{2}=|\phi_{0}\rangle_{\mathrm{H}}\langle\phi_{0}|\mathcal{V}_{\tau}^{(0)}+|\phi_{1}\rangle_{\mathrm{H}}\langle\phi_{1}|\mathcal{V}_{\tau}^{(1)}$, where the mismatched control basis are given by $|\phi_{0}\rangle_{\mathrm{H}}=\cos(\varphi/2)|0\rangle_{\mathrm{H}}-i\sin(\varphi/2)|1\rangle_{\mathrm{H}}$
and $|\phi_{1}\rangle_{\mathrm{H}}$ its orthogonal complement; $\mathcal{V}_{\tau}^{(0)}=e^{-i\pi\sigma_{y}^{C}/4}e^{-i\gamma\sigma_{x}^{C}/2}$
and $\mathcal{V}_{\tau}^{(1)}=\mathcal{V}_{\tau}^{0}\sigma_{x}^{C}$
are the feedback operations applied on the carbon nucleus and $\gamma$
is a suitable angle. As a result, all of the information-theoretic
quantities in the rhs of Eq.~(\ref{eq:4}) were obtained by performing quantum
state tomography (QST) \cite{key-22} along the experimental execution of the demon
protocol. Fig.~\ref{MD_FB_3} displays the entropy production in the experiment's
feedback-controlled operation. Negative values were obtained demonstrating
the realization of entropy rectification whose effectiveness worsens
as the bases mismatch increases. 

\begin{figure}[h]
\centering
	\includegraphics[scale=0.75]{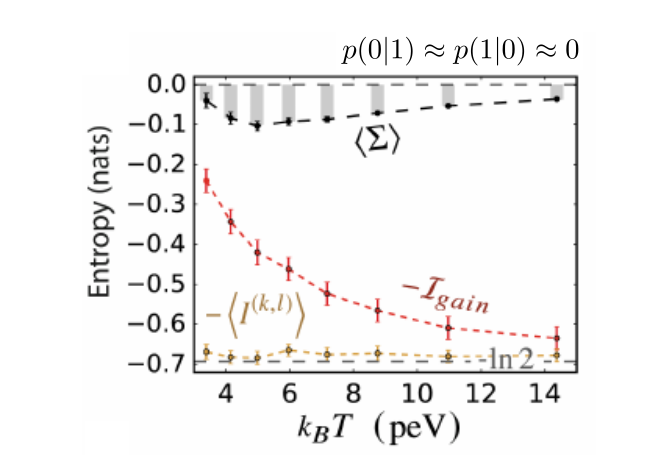}
	\caption{Information quantities and experimental verification of the bound imposed by means of the Eq.~(\ref{eq:4}). Entropy production (black), information gain bound, $\langle\Sigma\rangle\geq-\mathcal{I}_{gain}$ (red), 
	and mutual information bound $\langle\Sigma\rangle\geq-\langle I\rangle$ (dark yellow). Figure adapted from Ref.~\cite{Camati_2016}.}
	\label{MD_FB_4a}
	\label{MD_FB_4b}
\end{figure}

In Fig.~\ref{MD_FB_4a}, the authors noted that the bounds based on mutual information and information gain are not tight in a quantum scenario, as also anticipated by Eq.~(\ref{eq:4}). For this protocol, it was possible to show that $\langle I^{(k,\ell)}\rangle\geq\mathcal{I}_{gain}$ \cite{Camati_2016}. Despite the 4.5\% residual error in the trace distance for the zero mismatch case, the mutual information (between the system and feedback mechanism) experimentally achieved
is very close to its limit, $\langle I\rangle=-\sum_{\ell}p(\ell)\ln p(\ell)=\ln2\,\mathrm{nats}$
(natural unit of information) \cite{Camati_2016}.

The experiment \cite{Camati_2016} investigated the effect of several information qualities in the quantum version of Maxwell's demon. It also illustrates that the irreversibility of quantum nonequilibrium dynamics may be mitigated by assessing microscopic information, which is the effect of Maxwell's demon at work. A similar technique proposed in this work may be used for generic information-to-energy conversion protocols, such as information-based work extraction.

\subsection{Implementation of a Szilard's engine}\label{Szillard section}

\begin{figure}[h]
\centering
	\includegraphics[scale=0.4]{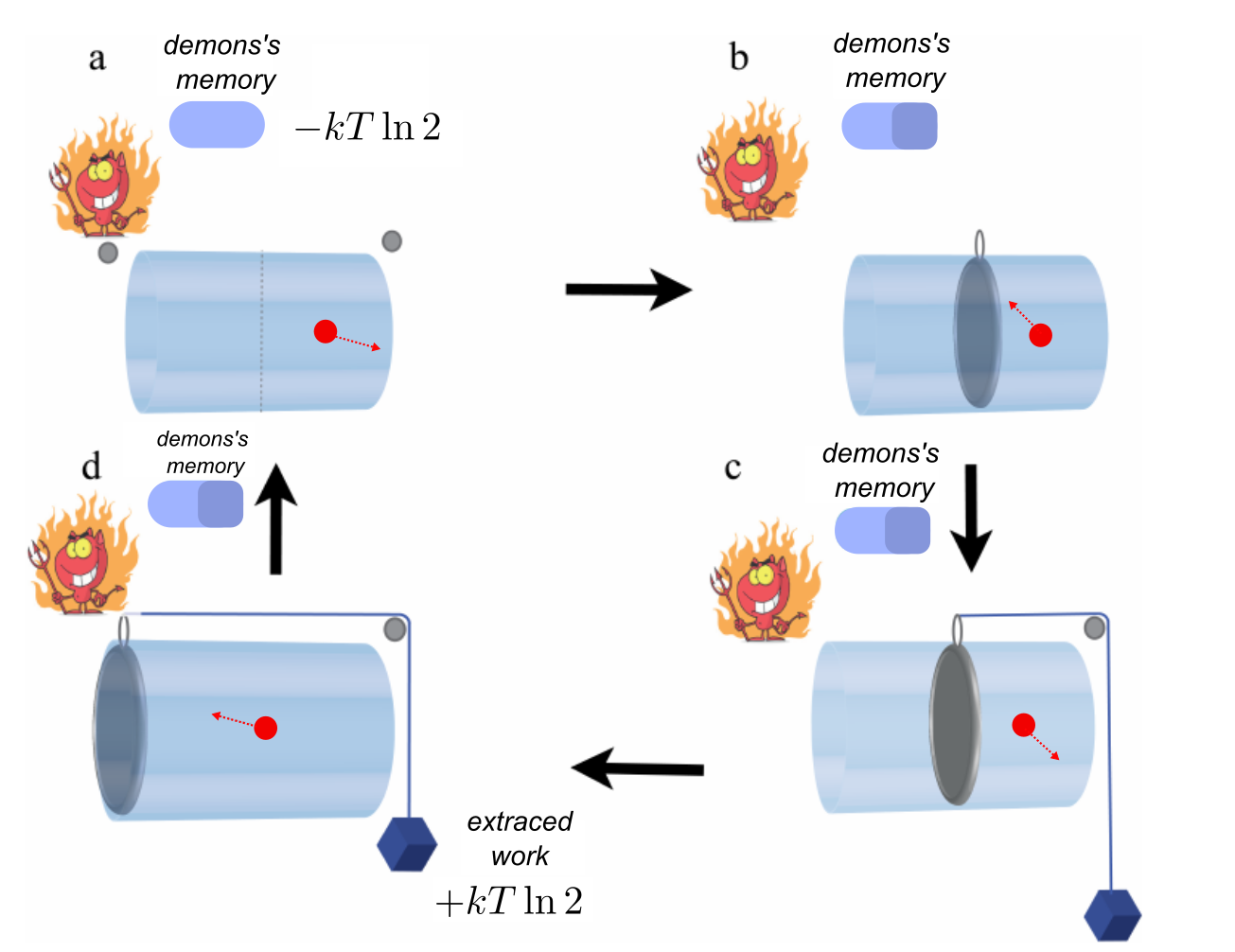}
	\caption{Classical Szilard's engine. The engine is composed of a container with only one particle, a thermal reservoir at temperature $T$, the demon's memory, and a weight. (a) to (b): Maxwell's demon determines that the particle is on the right side of the box and stores the result in a bit of memory. (b) to (c): the demon compresses the left side of the box at no cost given that there are no molecules on that side. (c) to (d): the demon allows the right side to expand until it fills the entire box, obtaining work equal to $k_{B} T \ln 2$, which culminates in lifting a weight. To complete the cycle, the demon must erase its memory from (c) to (d). According to the Landauer principle, this results in an energy cost per bit of $k_{B} T \ln 2$, see more in the next subsection \ref{Landauersection}.}
	\label{Szilard_01}
\end{figure}

In 1929, Szilard reformulated Maxwell's conundrum in which an intelligent being (the demon) may apparently violate the second law of thermodynamics using information about the state of a system \cite{Szilard,Szilard_2}. Using a classical engine, he considered a closed box with a single particle rather than multiple particles, a demon that measures the particle position and stores the information in its memory, a thermal reservoir at temperature $T$ from which heat is extracted, and a weight that is lifted using the work produced during the cycle. The Szilard's engine cycle is illustrated in Fig.~\ref{Szilard_01}. This thought experiment takes use of a compartment in the center of this box that functions as a piston. This compartment splits the box into two parts and the particle can be placed in either the left or right box using this arrangement. The demon observes whether the molecule is on the left or right side, stores this information in his memory, and introduces a division that will work as a piston. At this point, the partition where the particle is has half of the original volume. Because the molecule was not contacted during this process, no energy exchange occur, and so no work was performed. The molecule is then allowed to collide with the piston, which causes an isothermal expansion to the original volume (Fig.~\ref{Szilard_01}). At constant temperature, the expansion of a single-particle ideal gas against a piston produces a quantity of work $W=k_{B}T\ln2$, when the $V_{f}=2V_{i}$. The ultimate scenario involves a molecule occupying the whole space of the box, seemingly closing the cycle. Szilard's engine apparently shows a violation of Kelvin's statement of the second law, i.e., it is impossible to carry out a device operating cyclically where the only effect is absorbing energy in the form of heat from a single thermal reservoir and producing an equivalent amount of work~\cite{Balian_book,Maxwell_book}. 

Szilard's machine seemings to form a cycle that could be executed repetitively in order to produce large amounts of work. However, the cycle is not truly closed since the experimenter must record the measurement result in a classical memory. To return the system to its initial state and fully end the cycle, it must erase the record in its memory identifying which side the particle was on. This fact missed the attention of Szilard and other scientists at the time, and it was not until the 1960s that this issue was examined in greater depth for Landauer, as will be seen in the following. Despite several theoretical investigations for Szilard's engine in the quantum framework \cite{Ueda_2011,Kim_2012,Hai_2012,Liang_2012,Kim_PRL2013,Bengtsson_2013,Aydiner_2021}, few are experimental \cite{Peterson_arxiv,Spiecker}. This fact stems from the experimental challenges of successfully demonstrating that information might be used as fuel in quantum engines.

\begin{figure}[h]
\centering
	\includegraphics[scale=0.52]{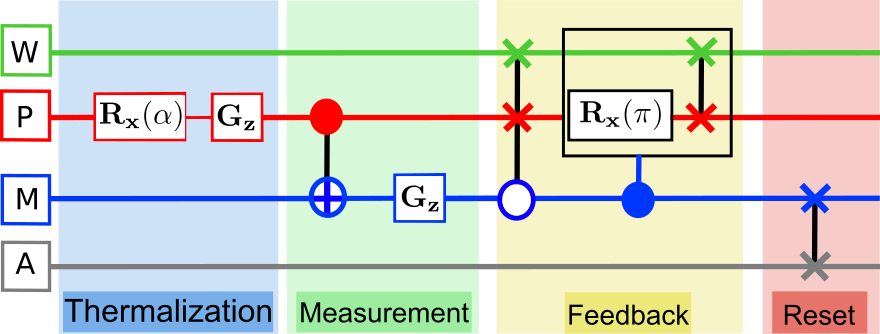}
	\caption{Quantum Szilard engine. The qubits W, P, M, and A represent the weight, the particle, the demon's memory, and the ancillary, respectively. $\mathrm{R_x(\alpha)}$ is a rotation by an angle $\alpha$ around the $x$-axis. $\mathrm{G_z}$ is a magnetic field gradient along $z$ direction that is applied to a system to remove the non-zero order coherence terms of its state. In this circuit there are three controlled gates \cite{Nielsen2011}; CNOT (in the measurement), a controlled SWAP (first gate of the feedback), and a composition of controlled rotation and Swap (second gate of the feedback). If the sphere in the control qubit is white, an operation will be done to the target qubits when the control qubit is excited; otherwise, the operation will be applied when the control qubit is in the ground state. A controlled gate will thereby modify just the states of the target qubits. The auxiliary and memory both begin the cycle in the same state. To reset the memory, a Swap gate is used to interchange the memory and auxiliary states. Figure adapted from Ref.~\cite{Peterson_arxiv}.}
	\label{Szilard_02}
\end{figure}

In the following, we describe the experiment reported in Ref.~\cite{Peterson_arxiv} consisting of an NMR implementation of a quantum Engine fuelled by information. The Szilard engine was realized in the quantum scenario employing a four-qubit system (four spin 1/2 nuclei). One qubit represents the weight, another the particle, a third the demon and its memory, and a fourth an auxiliary qubit used to erase the demon's memory. The quantum Szilard's engine cycle is depicted by means of the circuit in Fig.~\ref{Szilard_02}. This circuit is divided into four steps: an initial thermalization where the particle extracts heat from a reservoir on average; a measurement where the demon obtains the particle state information a feedback process in which information is used to fully convert heat into work and put the weight in its excited state; and an erasing process to complete the cycle.

The NMR technology was used to accomplish the quantum gates outlined in the quantum circuit Fig.~\ref{Szilard_02}. To experimentally represent the four qubits in Szilard's engine, the authors employed the four nuclear spins of the carbon atoms in the $\mathrm{^{13}C}$-labeled transcrotonic acid molecule \cite{Peterson_arxiv}. The state of the qubits can be controlled by rf pulses and magnetic field gradients and the magnetization of the nuclear spins may be measured \cite{key-22}. The sample is immersed in a strong and homogeneous magnetic field in the z-direction produced by a Bruker $700$ MHz spectrometer. In this experiment, the natural dynamics can be described, with a good approximation, by the following Hamiltonian
\begin{equation}
H_{0}=\sum_{i}\frac{\hbar\left(\omega_{i}-\omega_{r}\right)\sigma_{z_{i}}}{2}+\sum_{i\neq j}\frac{\pi\hbar J_{ij}\sigma_{z_{i}}\sigma_{z_{j}}}{4},
\end{equation}
where $\omega_i$ and $\sigma_{z_{i}}$ are the angular oscillation frequency and the Pauli matrix of the $i$-th nuclear spin, respectively. $\omega_r$ is the angular frequency of the rotating frame and $J_{ij}$ is the scalar coupling constant of the spins $i$ and $j$. In addition, the state of the qubits was controlled by radio-frequency pulses through the rf Hamiltonian $H_{RF}=\hbar\Omega(t)\sum_{i}({\cos\phi(t)\sigma_{x_{i}}+\sin\phi(t)\sigma_{y_{i}}})/{2}$, where $\Omega(t)$ and $\phi(t)$ are the pulse amplitude and phase, optimized to implement a quantum gate with high precision. Experimental values of all these quantities were reported in Ref.~\cite{Peterson_arxiv}.

\begin{figure}[th]
\centering
	\includegraphics[scale=0.36]{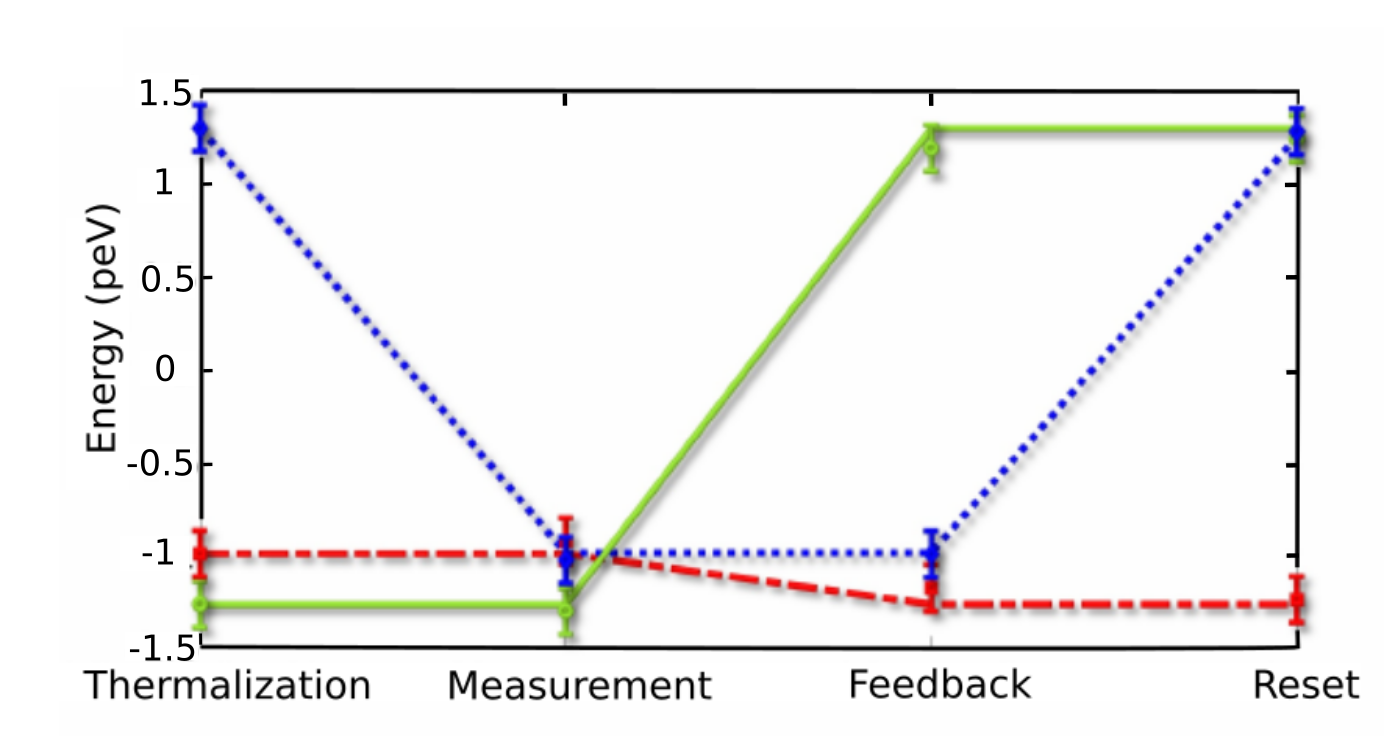}
	\caption{Experimental average energies in quantum Szilard's engine. The lines represent the theoretical prediction, while the points represent the experimental values of each system's average energy after each stage of the cycle. The particle, weight, and memory are represented by the colors red, green, and blue, respectively. The particle, weight, and memory Hamiltonian's are equal to $\omega\hbar\sigma_{z}$, with $\omega=2000$ rad/s. The particle was in the ground state before the thermalization with a reservoir at temperature $k_{B}T=1.33$ peV. Figure adapted from Ref.~\cite{Peterson_arxiv}.}
	\label{Szilard_03}
\end{figure}

The Szilard engine was implemented experimentally in four distinct configurations~\cite{Peterson_arxiv}. The weight and particle begin in their ground states in the first three, while the memory begins in its excited state. The temperature of the reservoir was modified in each implementation to demonstrate that the engine works at any temperature. The values used were $k_{B}T_{1}=1.33$ peV, $k_{B}T_{1}=2.51$ peV, and $k_{B}T_{1}=10.91$ peV \cite{Peterson_arxiv}. After each stage of the cycle, the average energy of the weight, particle, and memory were measured for each temperature configuration. The energy measured at each stage of the cycle agrees with the theoretical prediction, as shown in Fig.~\ref{Szilard_03}, where $k_{B}T=1.33$ peV. When we look at the error bars in Fig.~\ref{Szilard_03}, we can see that the weight is very near to its excited state after the feedback process, and the memory state has not changed considerably. Thus, the particle state information is employed to fully convert the heat extracted from the reservoir into work in order to put the weight in its excited state. Moreover, to confirm this full conversion, the authors also verify if the entropy variation of the weight is null during the feedback process. They used the weight state to calculate the entropy variation and obtained the result: $\Delta S_{T_1}=0.1$ peV, with an error of $\pm 0.2$ peV. Therefore, considering the error in the entropy and the amount of energy obtained by the weight ($\approx2.5$ peV), the amount of entropy produced due to experimental non-idealities was very small. To complete the cycle, the demon's memory is erased, and the average energy expended throughout this procedure equals $E_{T_1}=2.3\pm 0.1$ peV, which is the same amount of energy that decreased from the demon's memory during the measurement process, see Fig.~\ref{Szilard_03}. 

\begin{figure}[h]
\centering
	\includegraphics[scale=0.55]{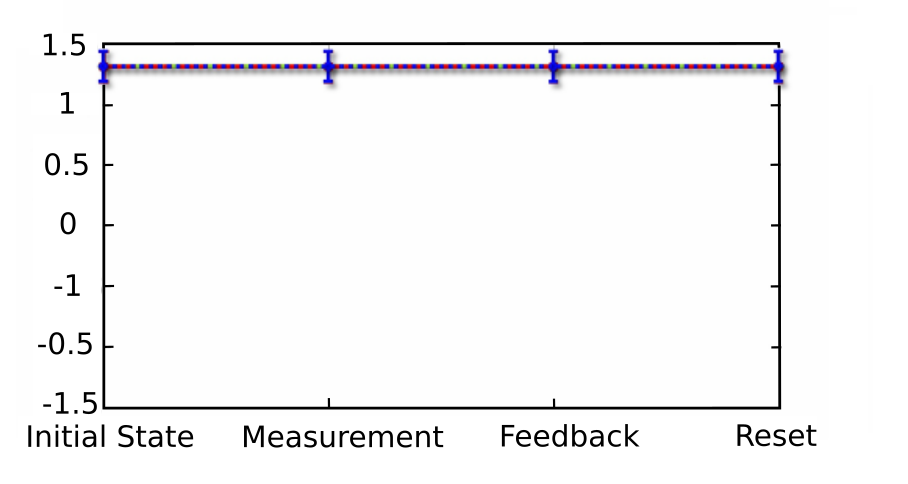}
	\caption{Experimental average energies in quantum Szilard's engine. The lines represent the theoretical prediction, while the points represent the experimental values of each system's average energy after each stage of the cycle. The particle, weight, and memory are represented by the colors red, green, and blue, respectively. All the systems start the cycle in the excited state and the thermalization is not performed. Figure adapted from Ref.~\cite{Peterson_arxiv}.}
	\label{Szilard_04}
\end{figure}

In the last implementation, they considered the case without thermalization, and the four qubits starting in their excited states. In this case, the states of the qubits should not change after the implementation of each step of the cycle. Then, as depicted in Fig.~\ref{Szilard_04}, it was possible to quantify how much energy is leaving the system (composed of the weight, particle, and memory) and going into the environment. As this amount of energy was close to zero, the system is isolated during the cycle. In other words, implementing the process with the whole system starting in its excited state the energy that leaves the system is very small \cite{Peterson_arxiv}. Aside from comparing the measured energy to the theoretical prediction, the authors used QST~\cite{key-22} to determine the state of each qubit at each stage of the cycle and compared it to the theoretical one using fidelity. Furthermore, the entropy fluctuation of the weight state was verified to demonstrate that the majority of the energy obtained by the weight represents work~\cite{Peterson_arxiv}.

To the best of our knowledge, this experiment was the first application of quantum Szilard's engine in the framework of NMR spectroscopy, effectively demonstrating that information may be employed as a fuel for single-reservoir engines.

\subsection{Information to energy conversion at the Landauer limit}\label{Landauersection}

As stated in the preceding subsection, Maxwell and Szilard believed that erasing memory in preparation for the next cycle would be a simple procedure and did not account for this in their analysis. Landauer discovered in 1962 that this is not a simple procedure. According to Landauer's principle, any irreversible computation produces inevitable entropy expressed as heat, which is dissipated to the computer. Landauer observed that this dissipated heat is bounded from below by the information theoretical entropy change \cite{Landauer_1,Landauer_2}. Therefore, the act of erasing memories must have an energy cost and this is precisely equivalent to the work that Maxwell's demon or Szilard's machine might acquire from the protocols outlined in the previous section. As a consequence, all of the work obtained must be used to erase the memory at the beginning of the next cycle, resulting in zero net work available for further purposes, as predicted by the second law. This discovery has been referred to as the ultimate exorcism of Maxwell's demon and it was thoroughly made by Bennett in 1982 \cite{Bennett,JDNorton_1,JDNorton_2}.

Landauer's principle is based on the idea that information erasure is an irreversible
process, with a fundamental heat cost associated with it. Thus, it provides an important connection between information theory and thermodynamics. Notwithstanding its crucial importance, only recently this principle has been experimentally performed involving individual classical or quantum systems \cite{Alexei,Ueda_Sagawa,Lutz,Pekola,Jun,LLYan,Celeri}. The main obstacle is the difficulty of doing single-particle experiments in the low dissipation regime where thermodynamic quantities such as heat, work, and entropy, behave as stochastic variables \cite{campisi, Denzler2021}. However, according to recent proposals, it is possible to obtain the quantum work distribution without having to perform these direct measurements by using phase estimation of an appropriately coupled ancilla, which explores the characteristic function of distribution and the thermodynamics quantities \cite{Dorner,Mazzola, serra2014}. 

According to Landauer's principle, the fact of erasure information produces an amount of heat $\langle Q_{\mathcal{R}}\rangle$ which is dissipated to the reservoir. Moreover, this average heat dissipated is related to the change in the von Neumann entropy of the system $\Delta {S}_{\mathcal{S}}={S}_{i}-{S}_{f}$, as follows
\begin{equation}\label{LandauerEq1}
    \langle Q_{\mathcal{R}}\rangle \geq k_{B}T\Delta {S}_\mathcal{S}.
\end{equation}
Here, $T$ is the temperature of the reservoir where the average amount of heat $\langle Q_{\mathcal{R}}\rangle$ is dissipated. Such a bound connects elements from two different systems and also restricts the heat absorbed by the bath to a quantity associated with the system's entropy change. Frequently, Landauer's principle is stated for binary systems in terms of the heating cost to erase one bit of information which in natural units (\textit{nats}) becomes $\langle Q_{\mathcal{R}}\rangle \geq k_{B} T\ln(2)$. This is a particular case of Eq.~(\ref{LandauerEq1}) for dichotomic variables. 

Recently, using the concepts of quantum statistical mechanics, Reeb and Wolf proposed a new version of Landauer's principle, which is described in terms of equality rather than inequality \cite{Reeb_Wolf} showing that Eq.~(\ref{LandauerEq1}) holds for processes satisfying the hypothesis: 

(I) the process involves a system $\mathcal{S}$ and a reservoir $\mathcal{R}$, both described by Hilbert spaces;

(II) the system $\mathcal{S}$ and the reservoir $\mathcal{R}$ are initially uncorrelated $\rho_{\mathcal{S}\mathcal{R}}=\rho_{\mathcal{S}}\otimes \rho_{\mathcal{R}}$;

(III) the reservoir $\mathcal{R}$ is initially in the Gibbs state $\rho_{\mathcal{R}}=e^{-\beta {H}_{\mathcal{R}}/\mathrm{tr}[e^{-\beta {H}_\mathcal{R}}]}$ with Hamiltonian ${H}_{\mathcal{R}}=\sum_{k}E_{k}\vert r_{k}\rangle\langle r_{k}\vert$, and inverse temperature $\beta=1/k_{B}T$;

(IV) the interaction between system $\mathcal{S}$ and reservoir $\mathcal{R}$ is unitary: $\rho_{\mathcal{S}\mathcal{R}}^{\prime}={U}\rho_{\mathcal{S}\mathcal{R}}{U}^{\dagger}$.

If these four conditions are satisfied, the quantum version of Landauer's principle can be described by the following equation
\begin{equation}\label{LandauerEq2}
    \frac{\langle Q_{\mathcal{R}}\rangle}{k_{B}T}=\Delta {S}_{\mathcal{S}} + {I}(\mathcal{S}^{\prime}:\mathcal{R}^{\prime})+{S}_{KL}(\rho_{\mathcal{R}}^{\prime}\vert\vert\rho_{\mathcal{R}}),
\end{equation}
where the symbol $\Delta {S}_{\mathcal{S}}$ denotes the change in von Neumann entropy of the system $\mathcal{S}$, the second term ${I}(\mathcal{S}^{\prime}:\mathcal{R}^{\prime})$ is the mutual information and quantifies the correlations built up between the system $\mathcal{S}$ and reservoir $\mathcal{R}$ during the process, and ${S}_{KL}(\rho_{\mathcal{R}}^{\prime}\vert\vert\rho_{\mathcal{R}})$ is the KL divergence which can be physically interpreted as the free energy increase in the reservoir. Relaxing any of these four criteria can lead to violations of Eq.~(\ref{LandauerEq1}). Thus, the processes that satisfy the four criteria above are known as Landauer processes \cite{Reeb_Wolf}. In consequence, the dynamics of the reservoir alone or the system are non-unitary. In this way, heat is produced in the reservoir as a consequence of modifying the entropy of the system. The changes in the entropy of the system $\mathcal{S}$ and the reservoir $\mathcal{R}$ can be computed by means of density operators $\rho_{S}=\mathrm{tr}_{\mathcal{R}}\left[\rho_{SR}^{\prime}\right]$ and $\rho_{\mathcal{R}}=\mathrm{tr}_{\mathcal{S}}\left[\rho_{\mathcal{S}\mathcal{R}}^{\prime}\right]$, while the average heat on the reservoir is computed by 
\begin{equation}\label{LandauerEq3}
    \langle Q\rangle=\mathrm{tr}\left[{H}_{\mathcal{R}}\left(\rho_{\mathcal{R}}^{\prime}-\rho_{\mathcal{R}}\right)\right].
\end{equation}

The heat absorbed by the reservoir is a stochastic variable \cite{Esposito,campisi}. Therefore, if before the unitary evolution $\mathcal{U}$, the reservoir had the probability $p_{m}=\exp\{-\beta E_{m}\}/Z_{\mathcal{R}}$ of was in the eigenstate $\vert r_{m}\rangle$ with energy $E_m$, after this process, the probability of finding the reservoir at the eigenstate $\vert r_{n}\rangle$, with energy $E_n$, is given by
\begin{equation}\label{LandauerEq4}
    p_{n\vert m}=\mathrm{tr}\left[{U}\vert r_{m}\rangle\langle r_{m}\vert\otimes\rho_{S}{U}^{\dagger}\vert r_{n}\rangle\langle r_{n}\vert\right].
\end{equation}
Therefore, with the probability $p_{m}p_{n\vert m}$, the reservoir $\mathcal{R}$ generates an amount of heat equal to $E_n-E_m$. Moreover, these probabilities give us a distribution for the heat \cite{Goold_heat}
\begin{equation}\label{LandauerEq5}
   P(Q)=\sum_{m,n}p_{m}p_{n\vert m}\delta\left[Q-\Delta E\right],
\end{equation}
where $\Delta E = E_{n}-E_{m} $ and the first moment of this probability distribution, $\langle Q\rangle=\int P(Q)QdQ$, is exactly the average heat showed in Eq.~(\ref{LandauerEq3}). 
Although it is often difficult to quantify the heat distribution Eq.~(\ref{LandauerEq4}) due to the invasive nature of projective measurements, we may measure its corresponding characteristic function $\mathcal{\chi}(t)$, which is computed by the Fourier transform
\begin{equation}\label{LandauerEq6}
   \mathcal{\chi}(t)=\sum_{m,n}p_{m}p_{n\vert m}e^{-i\Delta E t},
\end{equation}
which might also be written as \cite{Goold_heat,serra2014}
\begin{equation}\label{LandauerEq7}
   \mathcal{\chi}(t)=\mathrm{tr}\left[{U}\rho_{\mathcal{R}}\upsilon_{t}^{\dagger}\otimes\rho_{\mathcal{S}}{U}^{\dagger}\upsilon_{t}\right],
\end{equation}
where $\upsilon_{t}=e^{i{H}_{\mathcal{R}}t}$ is a unitary transformation on reservoir. It is important to emphasize that the average heat described by means of Eq.~(\ref{LandauerEq3}) corresponds to the first cumulant of the characteristic function $\mathcal{\chi}(t)$, Eq.~(\ref{LandauerEq7}).

In the NMR experiment, reported in Ref.~\cite{Celeri}, the authors study information to energy conversion at the Lander's principle in the quantum regime. Specifically, they considered a trifluoroiodoethylene $\mathrm{(C_{2}F_{3}I)}$ molecules, whose three $\mathrm{^{19}F}$ nuclear spins, with spin-$1/2$, represent the system, the reservoir, and the ancilla to measure the heat dissipated during the implementation of a global system-reservoir unitary interaction that changes the information content of the system. In fact, this work experimentally demonstrates the information to energy conversion in a quantum system operating at the Landauer limit.

\begin{figure}[h]
\centering
	\includegraphics[scale=0.92]{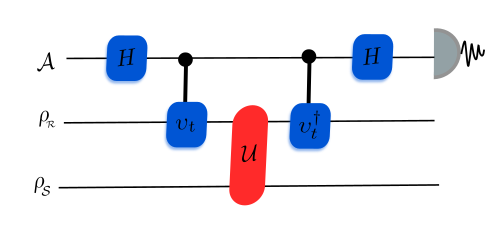}
	\caption{Quantum circuit used to determine the characteristic function $\mathcal{\chi}(t)$, Eq.~(\ref{LandauerEq7}). The auxiliary qubit ($\mathcal{A}$) is represented on the first line of the circuit, on the second line the reservoir ($\mathcal{R}$) and the third line represents the system ($\mathcal{S}$). In this circuit, we have used the unitary transformation $\upsilon_t=e^{i{H}_{\mathcal{R}}t}$. Figure adapted from Ref.~\cite{Celeri}.}
	\label{Landauer_quantum_circuit}
\end{figure}

To determine the heat characteristic function, without carrying out projective measurements, they used an auxiliary qubit $\mathcal{A}$ with must be initially prepared in the state $\vert +\rangle = [\vert 0\rangle+\vert 1\rangle]/\sqrt{2}$ and two logic gates, controlled by the auxiliary qubit, applied in the reservoir $\mathcal{R}$, as depicted in Fig.~\ref{Landauer_quantum_circuit}. 
The first controlled gate is applied before the unit evolution ${U}$ and implements the unit evolution $\upsilon_t$ in the state of reservoir $\mathcal{R}$ if the auxiliary qubit is in the $\vert 1\rangle$ state. In the same way, the second controlled gate is applied after the unit evolution ${U}$ and implements the unit evolution $\upsilon^{\dagger}_t$ in the state of the reservoir if the auxiliary qubit is in the $\vert 1\rangle$ state. In turn, following the quantum circuit, Fig.~\ref{Landauer_quantum_circuit}, is possible to find the characteristic function as following
\begin{equation}
    \mathcal{\chi}(t)=\mathrm{tr}\left[\left(\sigma_{x}-i\sigma_{y}\right)\rho_{\mathcal{A}}^{(3)}\right],
\end{equation}
where $\sigma_x$ and $\sigma_y$ are Pauli operators on the space of the ancilla and $\rho_{\mathcal{A}}^{(3)}$ is the final state of the auxiliary qubit after tracing out the state of the system $\rho_{\mathcal{S}}$ and reservoir $\rho_{\mathcal{R}}$ \cite{Celeri,Goold_heat}. Once the characteristic function is known, it is possible to take its inverse Fourier transform and determine the probability distribution of the heat $P(Q)$ and its average heat $\langle Q\rangle$ similar to what was done in \cite{serra2014}.

After a specific sequence of pulses, the three $\mathrm{^{19}F}$ nuclei was prepared in a state equivalent to 
\begin{equation}
    \rho=\vert+\rangle\langle+\vert\otimes\rho_{\mathcal{R}}\otimes\frac{\mathbb{I}_{\mathcal{S}}}{2},
\end{equation}
where the ancilla qubit $\rho_{\mathcal{A}}$, is prepared in state $\vert+\rangle$ and the reservoir qubit is prepared in the state
\begin{equation}\label{eq35}
    \rho_{\mathcal{R}}=\left(\begin{array}{cc}
\cos^{2}(\frac{\alpha}{2}) & 0\\
0 & \sin^{2}(\frac{\alpha}{2})
\end{array}\right).
\end{equation}
The symbol $\alpha$ represents the rotation angle present in the preparation pulse sequence~\cite{Celeri}. 

Comparing the Eq.~(\ref{eq35}) with the definition of the density matrix of a system in thermal equilibrium at finite inverse temperature $\beta$, it is possible to obtain a relation between the temperature and the rotation angle $\alpha$ with the reservoir temperature
\begin{equation}
    \beta^{-1} = \frac{2\pi\hbar J_{\mathcal{RA}}}{\log\left[\tan^{2}(\alpha/2)\right]},
\end{equation}
where $J_{\mathcal{RA}}$ is the scalar coupling between ancilla and reservoir qubit in the sample of trifluoroiodoethylene $\mathrm{(C_{2}F_{3}I)}$ molecules. Therefore, using this parameterization, it was possible to vary the angle $\alpha$ with the aim to produce different initial states for the reservoir \cite{Celeri}. The initial state of the system $\mathcal{S}$ was chosen to be maximally mixed, which represents the situation in which the system contains one bit of information, and consequently works as a memory. A different choice for the initial state of the system can result in a different quantity of average heat dissipated. Nevertheless, the validity of Landauer's principle is independent of this choice. 

The main purpose of this experiment was to estimate the heat dissipated by an elementary quantum logic gate at the ultimate limit, set by Landauer's principle. In this way, using rf pulses, two unitary operations were implemented: (i) the controlled-NOT or CNOT gate and (ii) the partial SWAP gate. The validity of Landauer's principle is completely independent of the choice of these particular operations.

\begin{figure}[h]
\centering
	\includegraphics[scale=0.3]{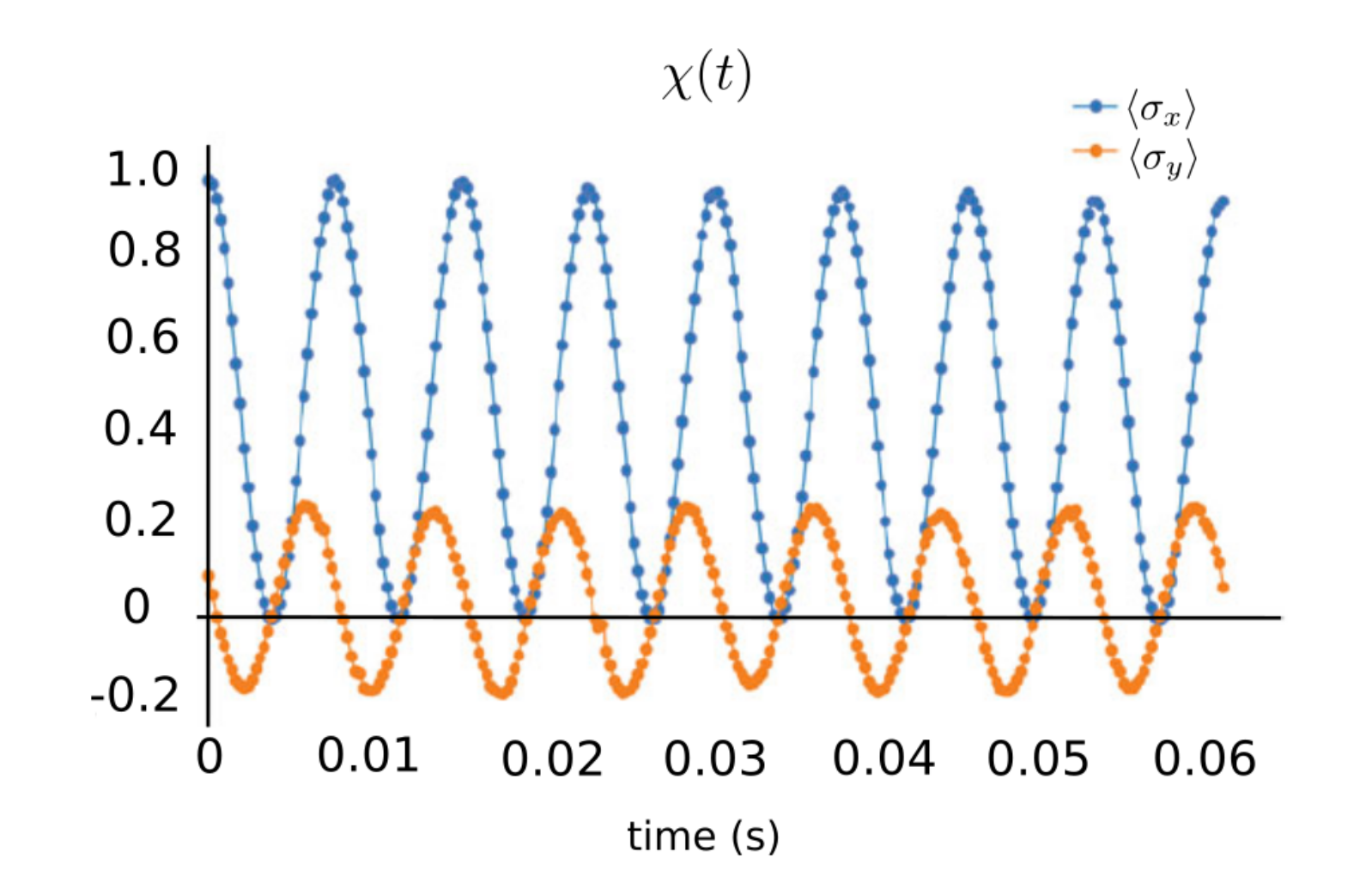}
	\caption{The characteristic function corresponds to heat distribution, Eq.~(\ref{LandauerEq5}), considering the unitary interaction between the system and the reservoir through the CNOT gate. The characteristic function, $\chi(t)$, was acquired by measuring the average values of the Pauli operators $\sigma_x$ and $\sigma_y$ on the space of the ancilla qubit, see Eq.~(\ref{LandauerEq7}). Figure adapted from the Ref.~\cite{Celeri}.}
	\label{characteristic_function}
\end{figure}

\begin{figure}[h]
\centering
	\includegraphics[scale=0.27]{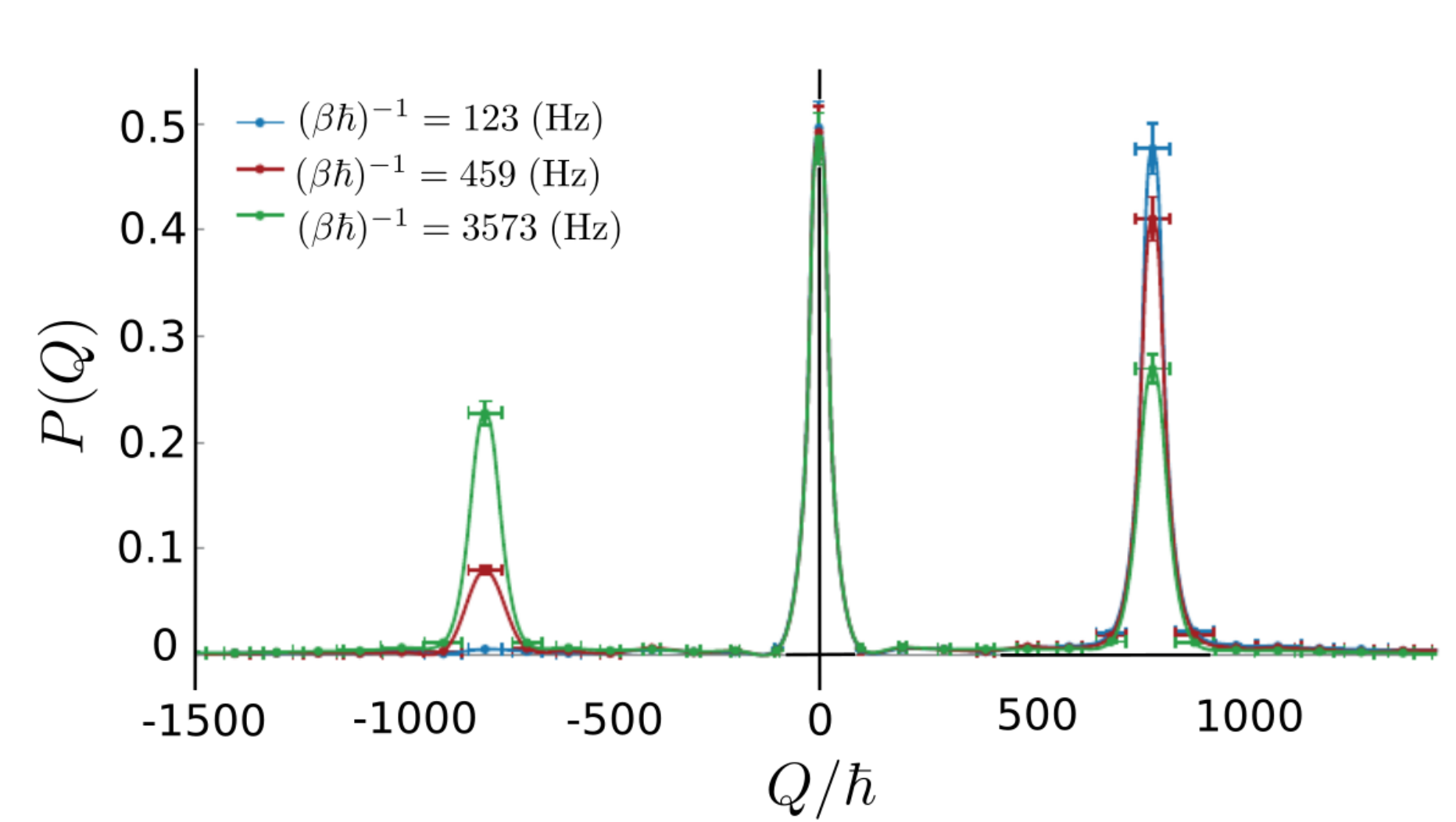}
	\caption{The heat probability distribution was acquired from the discrete Fourier transforming the characteristic function. This case shows the heat probability after the application of the CNOT gate and preparing the reservoir $\mathcal{R}$ at different temperatures. Figure adapted from the Ref.~\cite{Celeri}.}
	\label{heat_distribution}
\end{figure}

In case (i), they performed several experiments where the CNOT gate was taken as the unitary interaction between the system and the reservoir. It was assumed $\mathcal{S}$ is the control qubit and $\mathcal{R}$ is the target qubit. This unitary quantum gate was implemented by means of the following pulse sequence 
\begin{equation}
    {U}_{\mathrm{CNOT}}=R_{x}^{\mathcal{R}}\left(\frac{\pi}{2}\right)\mathcal{U}_{0}\left(\frac{\pi}{2}\right)R_{x}^{\mathcal{\mathcal{A}}}\left(\pi\right)\mathcal{U}_{0}\left(\frac{\pi}{2}\right)R_{y}^{\mathcal{\mathcal{R}}}\left(\frac{\pi}{2}\right),
\end{equation}
where $R_{k}^{i}\left(\theta\right)$ is a rotation on the $i$th qubit about direction $k$ by an angle $\theta$, whereas $\mathcal{U}_{0}$ represent a free evolution, i.e. generated by Hamiltonian without the radiofrequency part \cite{Celeri}. In this approach, the temperature of the reservoir was varied changing the parameter $\alpha$, as aforementioned. Figure~\ref{characteristic_function} depicts an example of the experimental characteristic function, whereas in Fig.~\ref{heat_distribution} we can see the heat distribution at different values of inverse temperature $\beta$. The central peak at $Q = 0$, in the heat distribution, corresponds to the cases where the energy eigenstate does not change, whereas $Q > 0$ means a transition from a low-energy state to a high-energy state has occurred, and $Q < 0$ represents the reverse situation. For this particular gate (CNOT), it is straightforward to see that the theoretical entropy change is $\Delta \mathcal{S} = 0$. However, as it is clearly shown, there are instances where $Q < 0$, seemingly in violation of the Landauer principle. Reinforcing the statistical concept of the second law, these events are fluctuations and the stochastic nature of the thermodynamic variables in this domain is highlighted. As we can see, Fig.~\ref{heat_distribution}, even though there is a chance to spot a momentary violation of Landauer's bound in the quantum domain, the average heat is greater than the entropy variation, supporting the idea that Landauer's principle as well as the second law are valid on average but may not always be verified to a one-shot experimental realization.

\begin{figure}[h]
\centering
	\includegraphics[scale=0.47]{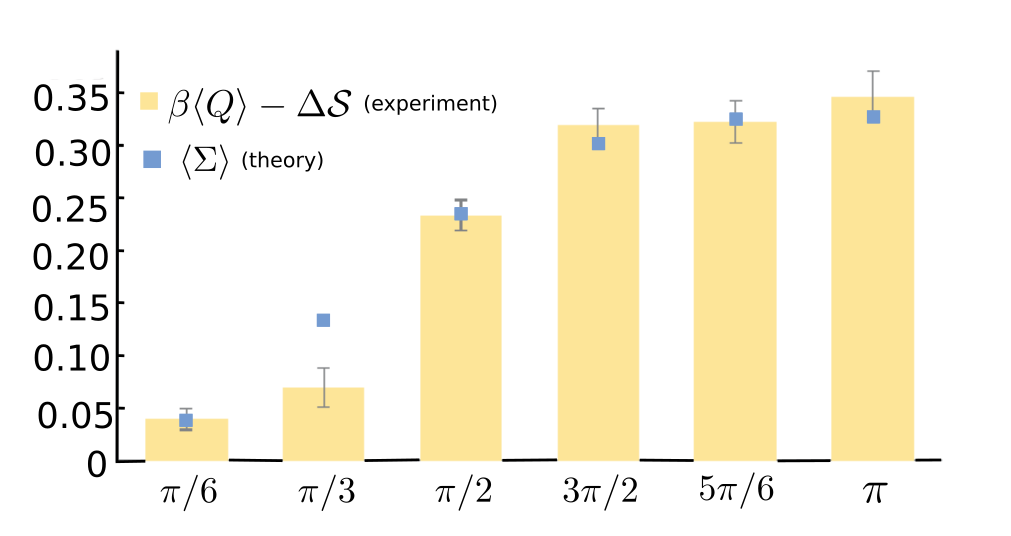}
	\caption{Landauer limit for partial SWAP gate. The experimentally measured gap between heat and entropy when compared with the theoretically computed irreversible entropy production. Figure adapted from the Ref.~\cite{Celeri}.}
	\label{entropy_production_Landauer}
\end{figure}

In case (ii), in order to reach the Landauer limit, they performed several experiments where the partial SWAP gate was taken as the unitary interaction between the system and the reservoir. Depending on the value of the tuned parameter $\varphi$, this unitary operation becomes a total or partial SWAP gate. The unitary operation with $\varphi = \pi$, illustrates the example of the full erasure protocol since the final state of the system $\mathcal{S}$ ends in the thermal state irrespective of its initial state. Although the parameter $\varphi$ may vary, in all cases Landauer principle holds on \cite{Celeri}. Owing to the finite size of the thermal reservoir $\mathcal{R}$, this process generates a quantity of average entropy production, $\langle\Sigma\rangle$, that can be computed as $\langle\Sigma\rangle = \beta\langle Q\rangle-\Delta\mathcal{S}$ \cite{Esposito2010}. In Fig.~\ref{entropy_production_Landauer}, the authors plotted the experimentally measured $\langle\Sigma\rangle$ along with the theoretically computed quantity. The agreement between experiment and theory confirmed that it is possible to measure the heat dissipated by an elementary quantum logic gate at the ultimate limit, set by Landauer's principle \cite{Celeri}.

%

\subsection{Reversion of heat flow using non-classical correlations}\label{sec:thermodynamic_arrow_of_time}

Clausius's statement of the second law of thermodynamics says that heat flows from a hot body to a cold one spontaneously \cite{Clausius1879}. Furthermore, Boltzmann associates this irreversible behavior to the initial conditions of the microscopic dynamics in the system \cite{Boltzmann1877,Lebowitz1993, Zeh2007}. However, initial conditions not only induce irreversible heat flow but also determine the direction of the heat current. Besides that, the basic assumption of classical thermodynamics is that systems are uncorrelated before thermal contact. However, Refs. \cite{Partovi2008, Jevtic2012, Jennings2010, Bera2017} theoretically suggest that for quantum correlated systems heat can flow from the cold to the hot body without any external action. Recently, NMR techniques have been employed to experimentally access this quantum behavior \cite{Micadei2019}. It was considered two nuclear spins-1/2, in the $\mathrm{^{13}C}$ and $\mathrm{^{1}H}$ nuclei of a $\mathrm{^{13}C}$-labeled CHCl$_3$ liquid sample diluted in Acetone-D6, to investigate heat transfer at the quantum domain. The heat exchange is studied in a time interval of a few milliseconds which is much shorter than any relevant decoherence time of the system. Also, the dynamics of the two spins in the sample are assumed to be basically isolated and consequently, the energy is conserved in each sample molecule. 

\begin{figure}[h]
\centering
	\includegraphics[scale=0.14]{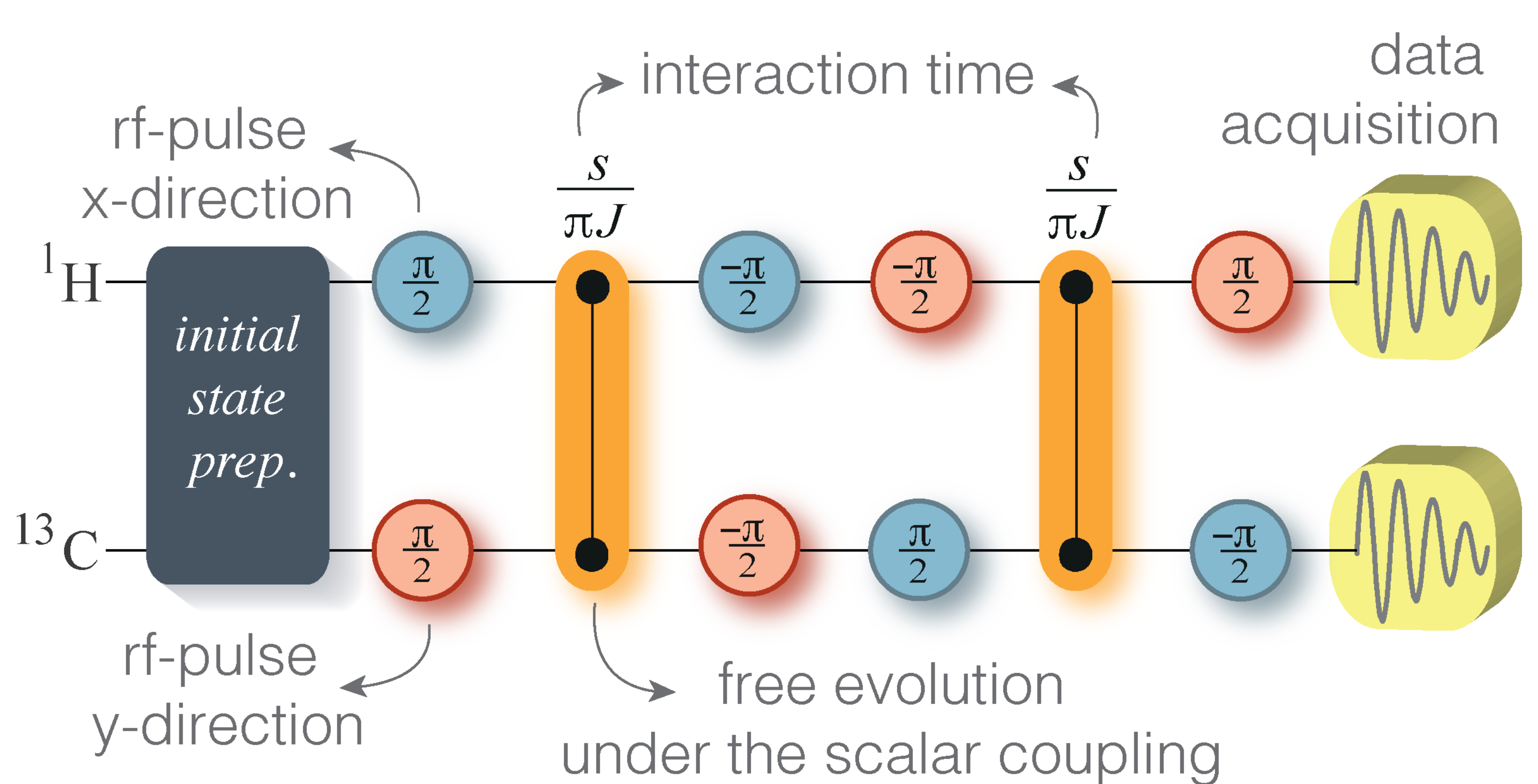}
	\caption{Experimental rf pulse sequence for the partial thermalization process. The blue (black) circle represents $x(y)$ rotations by the indicated angle. The orange connections represent a free evolution under the scalar coupling, ${H}_J = (\pi\hbar/4) J\sigma_z^H \sigma_z^C$, between
the 1H and 13C nuclear spins during the time indicated above the symbol. We have performed 22 samplings of the interaction
time $\tau$ in the interval 0 to 2.32ms. Figure adapted from Ref. \cite{Micadei2019}.}
	\label{Reversion_figure}
\end{figure}

In order to study the heat flow between the two nuclear spins initially correlated, the system is prepared to have an initial state equivalent to
\begin{equation}
\rho_{AB}^0 = \rho_A^0\otimes\rho_B^0 + \xi_{AB},
\end{equation}
with $\xi_{AB} = \alpha |01\rangle\langle 10| + \alpha^{\star} |10\rangle\langle01|$ being the correlation term and $\rho_i^0 = \text{exp}\left(-\beta_i {H}_i)/Z_i\right)$ the local thermal state, $i = A, B$. The state $|0\rangle$ and $|1\rangle$ represent the ground and excited eigenstates of the Hamiltonian ${H}_i$. In order to guarantee the positivity, the condition $\alpha \leq \text{exp} \left[-h \nu_0 (\beta_A + \beta_B)/2\right]/(Z_A Z_B)$, with the frequency $\nu = 1$ kHz and ${H}_i = h\nu_0(\mathbb{I} - \sigma_z^i)/2$. Effectively, the partial thermalization is implemented by the effective interaction Hamiltonian, ${H}^{\text{eff}}_{AB} = (\pi \hbar/2)J(\sigma^A_x \sigma^B_y - \sigma^A_y \sigma^B_x)$, $J = 215.1$Hz, such that the evolution operator is ${U}_\tau = \exp(-i \tau {H}^{eff}/\hbar)$. Such an effective evolution is produced by the pulse sequence depicted in Fig.~\ref{Reversion_figure}.

To perform the experiment reported in \cite{Micadei2019}, it is noted that the total Hamiltonian commutes with the interaction Hamiltonian, such that the thermalization operation does not involve any work. All the energy exchanged between the two spins is characterized as heat. The heat flow at any time $\tau$, in aforementioned scenario can be written as~\cite{Micadei2019}
\begin{equation}
\Delta\beta Q_\text{B}=\Delta I(A{:}B)+S_{\text{KL}}(\rho_\text{A}^{\tau}||\rho^0_\text{A})+S_{\text{KL}}(\rho_\text{B}^{\tau}||\rho^0_\text{B}),\label{eq:5}
\end{equation}
where $\Delta I(A:B)$ the change of the mutual information between the two spins, $I(A:B) = S_A + S_B - S_{AB}$, $\Delta\beta=\beta_\text{B}-\beta_{A}\geq0$ and $S_{\text{KL}}(\rho_{i}^{\tau}||\rho_{i})=\mathrm{tr}_{i}\,\rho_{i}^{\tau}(\ln\rho_{i}^{\tau}-\ln\rho_{i})\geq0$
denotes the relative entropy between the evolved $\rho_\text{A(B)}^{\tau}=\mathrm{tr}_\text{B(A)}\mathcal{U}_{\tau}\rho_\text{AB}^{0}\mathcal{U}_{\tau}^{\dagger}$
and the initial $\rho_\text{A(B)}^{0}$ reduced states. The latter quantifies
are related to the entropy production associated with irreversible heat transfer. From Eq.~(\ref{eq:5}), we observe that the
direction of the {energy current} is therefore reversed whenever the decrease
of mutual information compensates the entropy production. In the experiment reported in Ref.~\cite{Micadei2019}, the two-qubit system was initially prepared with effective spin temperatures $\beta_A^{-1} = 4.66 \pm 0.13$peV $(\beta_A^{-1} = 4.30 \pm 0.11)$ peV and $\beta_B^{-1} = 3.31 \pm 0.08$ peV $(\beta_B^{-1} = 3.66 \pm 0.09)$ peV for the uncorrelated (correlated) case $\alpha = 0.00 \pm 0.01$ $(\alpha = -0.19 \pm 0.01)$ \cite{Micadei2019}. 

\begin{figure}[h]
\centering
	\includegraphics[scale=0.47]{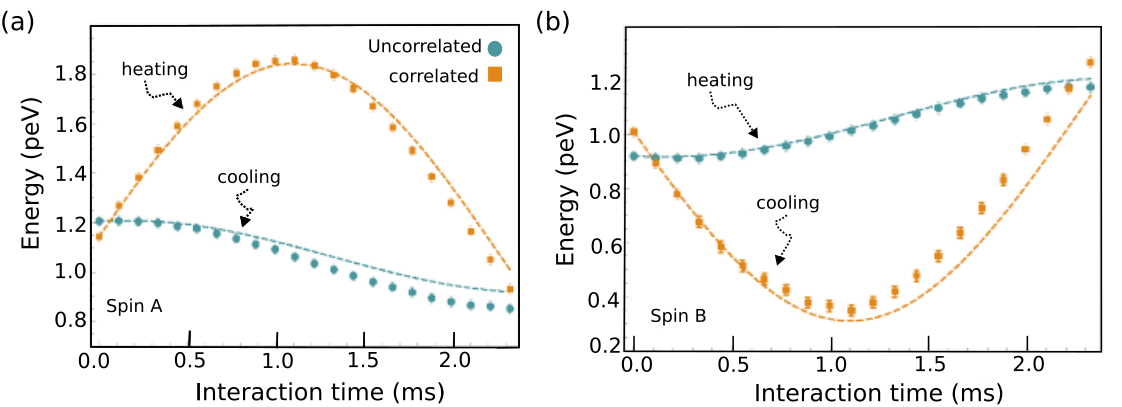}
	\caption{Dynamics of heat. \textbf{(a)} Internal energy of spin A along the partial
thermalization process. \textbf{(b)} Internal energy of spin B. In the absence of initial correlations, the hot spin A cools down and
the cold spin B heats up (cyan circles in panel \textbf{a} and \textbf{b}). By contrast, in the presence of initial quantum correlations, the
heat current is reversed as the hot spin A gains and the cold spin B loses energy (orange squares in panel \textbf{a} and \textbf{b}). This
reversal is made possible by a decrease in the mutual information and the geometric quantum discord. Figure adapted from the Ref. \cite{Micadei2019}.}
	\label{Reversion_figure02}
\end{figure}

Figure \ref{Reversion_figure02} shows the experimental result of the reverse in the heat flow using quantum correlations implemented using NMR techniques \cite{Micadei2019}. It is observed the reversal of the heat flow between the two quantum-correlated spins with different effective temperatures, i.e., heat flows from the colder spin to the hotter spin. During this reversal behavior, it is also measured the decrease of the initial quantum correlations during the dynamics~\cite{Micadei2019}. This highlights that depending on the initial conditions, uncorrelated or correlated in the experiment, the heat flow direction can be reversed. This experiment contributes to showing the high degree of control in exploring the quantum nature of the two-spin system as well as carrying out a test-of-principle experiment about the role played by non-classical correlations in quantum thermodynamics.

\section{Conclusion and outlook}\label{Conclusion and outlook}

We presented a short overview of some approaches to quantum thermodynamics, highlighting experimental implementations using the NMR platform. One of the biggest motivations of quantum thermodynamics has been to extend thermodynamic principles to the quantum world, where energy fluctuations coherence, and quantum correlations take into play. Addressing these quantum correlations has made it possible to discuss the role of quantum resources in devices such as thermal engines, refrigerators, thermal accelerators, and batteries. 

Quantum stochastic thermodynamics has brought the possibility to deal with fluctuation theorems enabling us to consider explicitly energy fluctuations in non-equilibrium dynamics and settle relations that connect equilibrium properties of thermodynamics relevance with non-equilibrium features, beyond dealing with processes involving measurement and feedback control. Once fluctuations have a pivotal role in small systems, many efforts have been made to establish a trade-off between fluctuations and dissipation in stochastic thermodynamics. A set of relations addressing this goal has been collectively referred to as thermodynamic uncertainty relations (TURs) that can constrain power, fluctuations, and the entropic cost of quantum tasks. Additionally, investigating the connection between information and energy quantum information thermodynamics has aided to solve misinterpretations in famous conundrums of physics, such as Maxwell's demon and Szilard's engine besides addressing the thermodynamic cost of quantum operations.

This vibrant branch of research is tremendously active and provides the intersection of several areas, including stochastic thermodynamics, quantum physics, and condensed matter, beyond having meaningful connections with computing science and engineering. Owing to the interdisciplinary and broad nature of this field, many financial efforts have been made to develop quantum thermodynamic devices and speed up this new industrial revolution on the tiniest scale (quantum technology 2.0). Although we have focused on the NMR technique, which significantly contributed to the field due to its precision in measurement and control, quantum thermodynamics has been widely fueled by experiments in several physical approaches such as trapped ions, cold atomic, quantum optics, single electron transistors, cavity QED, Nitrogen-Vacancy centers, and superconducting circuits. 

\section*{Acknowledgments}
The authors acknowledge the funding agencies CAPES (Grants No. 88887.354951/2019-00 and No. 88887.499539/2020-00), FAPESP (Grant No. 2019/04184-5), CNPq (Grants No. 310430/2018-6 and No. 142531/2020-0), and the National Institute for Science and Technology of Quantum Information (CNPq, Grant No. INCT-IQ 465469/2014-0). The authors are grateful to the Multiuser Central Facilities (CEM-UFABC) for the experimental support. P.R.D. acknowledges support by the Foundation for Polish Science (FNP) (IRAP project, ICTQT, Contract No. MAB/2018/5, co-financed by EU within Smart Growth Operational Programme). R.M.S. also acknowledges Ministry of Science and Technology of China, through the High-End Foreign Expert Program (Grant No. G2021016021L).

\vspace{2mm}
{\noindent \large \textbf{{Author contributions}}} \\
All authors discussed the topics and contributed to write the paper.

\vspace{2mm}
{\noindent \textbf{\large{Competing interests}}}\\  
The authors declare no competing interests.

\end{document}